\documentclass[aps,prd,preprint,nofootinbib,groupedaddress]{revtex4-1}
\usepackage{amssymb}
\usepackage{amsmath}
\usepackage{amsfonts}
\usepackage{color}
\usepackage{graphicx}
\usepackage{bm}
\usepackage{braket}
\usepackage{enumerate}

\usepackage{amsthm}
\usepackage{here}

\newcommand{\sub}[1]{\ensuremath{_{\mathrm{#1}}}}
\newcommand{\sps}[1]{\ensuremath{^{\mathrm{#1}}}}
\newcommand{\rre}{\mathrm{Re}}
\newcommand{\rim}{\mathrm{Im}}
\newtheorem*{definition*}{Definition}

\begin{document}

\title{ Spatially Overlapped Partners \\
in Quantum Field Theory }
\author{Jose Trevison}
\author{Koji Yamaguchi}
\author{Masahiro Hotta }

\affiliation{Graduate School of Science, Tohoku University,\\ Sendai,
980-8578, Japan}

\begin{abstract}
In quantum field theory particles are physically defined as what
Unruh-DeWitt particle detectors observe. By detecting a particle mode $A$, a
reduced density operator for a quantum state of $A$ is constructed. Even if
the entire quantum state of the quantum field is pure, the state of $A$ is not
pure but mixed due to entanglement between other subsystems. The partner
mode $B$ of the field is defined as a purification partner of $A$ such that the
$AB$ system in a pure state. We show that, without any fine-tuning of the
particle detector design of $A$, the weighting function of partner $B$ has spatial
overlap of that of $A$. We show a general formula of partner $B$ associated with
arbitrarily fixed $A$ of a free field in a general Gaussian state. We
demonstrate an example of memory effects in an expanding
Freedman-Roberson-Walker universe.
\end{abstract}

\maketitle

\section{Introduction}

A quantum field is capable of playing a role of quantum information storage.
After a quantum operation dependent on unknown parameters is performed to
the field, the quantum state stores the memory of the parameters. In what
kind of form does the field keep the information? There exist a lot of
options. For instance, a two-body subsystem in a pure entangled state is
able to keep the information. The two-body system is referred to as an
entangled partner \cite{HSU}. Since a field in the vacuum state has an
infinite number of partners due to the ultraviolet divergence, it is well
known that the entanglement entropy diverges to infinity. By use of the huge
entanglement, a quantum field may attain large information capacity. From
this point of view, the entanglement of the partners can be expected to
provide relevant applications for future quantum information technology,
such as entanglement harvesting \cite{R1,R2}.

Besides, the notion of entangled partner has shed light on fundamental
physics like the black hole information loss problem \cite{H}. In \cite{HSU}, the partner mode corresponding to a Hawking mode of a free field is
explicitly identified. It turns out that the partner is a local zero-point
fluctuation of the field. This may avoid a serious flaw of the information
recovery scenario at the last burst of a black hole so as to maintain the
unitarity of the process. It is widely argued that evaporating black hole
energy of the order of the Planck scale is too small to emit the whole
inside information to outside \cite{P}. Since the amount of information is
not elementary particle size but astrophysical size, the information
carriers seem to request a huge number of highly excited states, and much
larger energy than the Planck energy. However, as pointed out in \cite{W}
and \cite{HSU}, the zero-point fluctuation emitted at the last burst is able
to retrieve the whole inside information because the fluctuation flow
requires zero energy cost.

In quantum field theory particles are physically defined as what
Unruh-DeWitt particle detectors observe \cite{U}, \cite{DW}. Measuring a
particle mode $A$ by the detectors is capable of identifying a reduced
density operator for a quantum state of $A$ via quantum state tomography
protocols. Even if the entire quantum state of the quantum field is pure,
the quantum state of a subsystem is not generally pure but becomes mixed due
to entanglement between other subsystems. The partner mode $B$ associated to 
$A$ in the field is defined as a purification partner of $A$ such that the $AB$ system is in a pure state. In \cite{HSU}, a special type of Unruh-De
Witt detector for a Hawking particle succeeded in capturing the parter of a
Hawking particle and clarifying its interesting properties. The mode of $A$
is fixed by operators consisting linear combination of a field operator and
its conjugate momentum operator with some weighting functions localized in a
spatial region. The partner mode $B$ associated with $A$ is determined in a
similar way by a linear combination of the field operator and its conjugate
momentum. The weighting functions of $B$ has no overlap of spatial support
with that of $A$. This means that the Hawking particle has a \textit{spatially separated partner} (SSP) in \cite{HSU}.

\begin{figure}[]
\centering
\includegraphics[width=8cm]{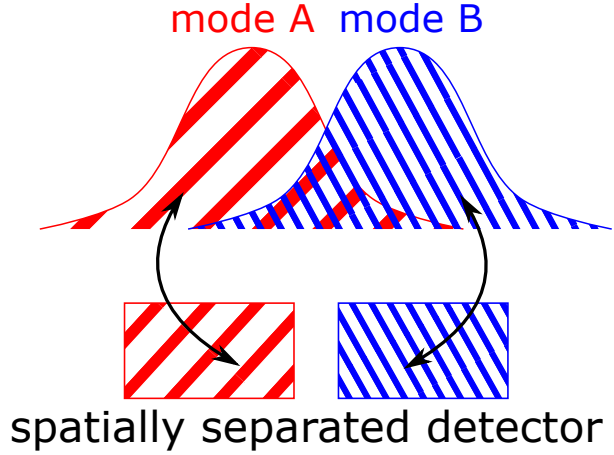}
\caption{A schematic picture of information extraction using spatially separated detectors. The rectangle with red (resp. blue) line pattern denotes a detector for mode $A$ (resp. mode $B$). Since they do not have any spatial overlap, it is difficult to extract whole quantum information imprinted in SOPs.\label{fig_detector1}}
\label{GaussianA}
\end{figure}

\begin{figure}[]
\centering
\includegraphics[width=8cm]{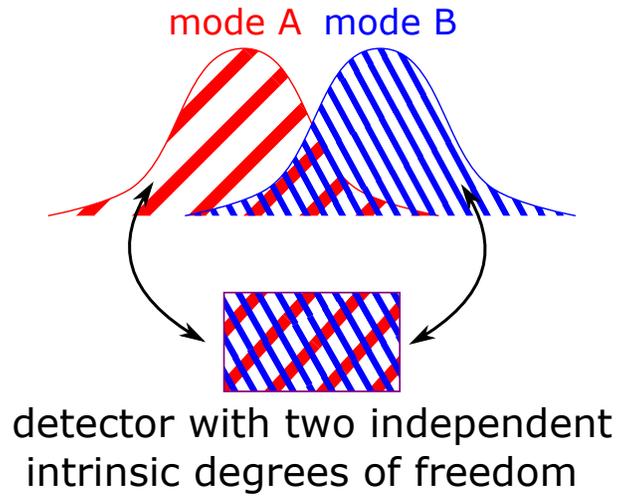}
\caption{A schematic picture of information extraction using a detector with two independent intrinsic degree of freedoms. The rectangle with red and blue line pattern denotes the detector. With such a special device, it is possible to extract whole quantum information imprinted in SOPs.\label{fig_detector2}}
\label{GaussianA}
\end{figure}

In this paper, we elaborate a more general class of partners of a free
scalar field in an arbitrary Gaussian state. We show a general formula of
partner $B$ associated with an arbitrarily fixed $A$ of a free field in a
general Gaussian state. It turns out that, without any fine-tuning of the
choice of $A$ mode, the spatial support of weighting functions of $B$ mode
has nonzero overlap with that of $A$ mode. This implies that a particle
observed by a general Unruh-De Witt detectors is accompanied by a \textit{spatially overlapped partner} (SOP) for purification of the particle.
Though the spatial overlap of $A$ and $B$ happens, it is possible to
consider quantum entanglement between $A$ and $B$ since the operators of
each system commute to each other and establish locality of $A$ and $B$ for
the definition of the entanglement. In the similar way of usual SSP cases in 
\cite{HSU}, the pure states of SOP are also able to play a role of quantum
memory devices about unknown parameters by imprinting them via
parameter-dependent dynamical processes. In order to demonstrate that
explicitly, we consider a simple example of SOP of a scalar field in an
expanding universe with an expansion rate parameter $\rho $. We show that there exists the $\rho $-dependence of entanglement entropy
between a localized particle mode $A$ and its SOP mode $B$. Such an analysis
of SOP may allow us to construct a more sensitive model for checking
cosmological Bell inequality breaking in cosmic microwave back ground \cite{m}. The aim of this paper is to stress a new concept of information storage by SOP in quantum field theory, which have not been pointed out to date. Though it is significant to analyze SOP information storage in black hole evaporation as well as SSP, it requires a more complicated calculation, that is outside of the reach of the present paper. It is also worthwhile to stress that a partner
exists for an arbitrarily fixed particle mode in a quantum field in a
general pure state, as will be mentioned in \ref{sec_definition}, and the
partner is expected to be an SOP in typical cases. Thus SOP may be applied
to a wide class of physics issues including the black hole information loss
problem. One might be afraid that the spatial overlap of $A$ and $B$
disturbs extraction of the imprinted information in SOP. As depicted in
Figure \ref{fig_detector1}, it is difficult to read out the information of SOP by using two
spatially separate detectors. However, by using a special quantum swapping
device as depicted in Figure \ref{fig_detector2} and possesses two independent intrinsic
degrees of freedom associated with $A$ and $B$, the information as well as
the entanglement can be extracted and read out perfectly \cite{TYH}.


In Section \ref{sec_definition}, we prove existence of partner $B$ for an
arbitrarily fixed mode $A$ of a field in a general state. In Section \ref{sec_partner_vac}, the partner formula for the vacuum state for a free
scalar field is derived. Without any fine-tuning, the partner becomes an
SOP. 
In Section \ref{sec_partner_general}, we derive a general expression for partner formula, which is applicable to any Gaussian state and any complete set of canonical operators. 
In Section \ref{sec_partner_CS}, we demonstrate how SOPs store
information about parameters of dynamical evolution of the field. As a
simple example, an expanding universe with an expansion rate parameter $\rho$ is considered. There actually exists $\rho $-dependence of entanglement
entropy between a localized particle mode and its SOP. In Section \ref{sec_SandD}, conclusions are presented.

Throughout this paper, scalar field theory is treated as the continuum limit of harmonic oscillator chain. We do not discuss any subtle problems regarding the continuum limit. Our results are applicable if the limit can be taken properly.

In this paper, the natural unit is adopted: $c=\hbar=1$.

\section{Correlation Function Definition of Purification Partner}

\label{sec_definition}

In this section, we start with a definition of a partner for an arbitrary
mode $A$ of a free quantum field in a general state $|\Psi \rangle $ by
using a Unruh-De Witt detector and measurable correlation functions of $A$.
This definition provides a significant advantage which provides direct
methods to verify the partner of $A$ by realistic physical experiments. The
mode $A$ is defined by what an extended Unruh-De Witt particle detector
observes. Let $\hat{\varphi}(\bm{x})$ and $\hat{\Pi}(\bm{x})$ be a free
scalar field and its conjugate momentum in a $(d+1)$-dimensional curved
spacetime. Let us consider a continuous-variable Unruh-De Witt detector with a
measurement interaction as 
\begin{equation*}
\hat{H}_{\mathrm{meas}}(t)=\lambda (t)f\left( \hat{q}_{A}(t),\hat{p}_{A}(t)\right) \hat{P}_{\mathrm{D}}(t),
\end{equation*}
where $\lambda (t)$ is a time-dependent coupling between the field and the
detector, and $\hat{P}_{\mathrm{D}}(t)$ is a
momentum operator conjugate to a pointer position operator $\hat{Q}_{\mathrm{
D}}(t)$ of the detector. Also $f\left( q,p\right) $ is a real function of $q$
and $p$, and $\hat{q}_{A}(t)$ and $\hat{p}_{A}(t)$ are Heisenberg operators
associated with linear combination of the field operators as 
\begin{align}
\hat{q}_{A}& =\int d^{d}\bm{x}\left( x_{A}(\bm{x})\hat{\varphi}(\bm{x})+y_{A}(\bm{x})\hat{\Pi}(\bm{x})\right) , \\
\hat{p}_{A}& =\int d^{d}\bm{x}\left( z_{A}(\bm{x})\hat{\varphi}(\bm{x})+w_{A}(\bm{x})\hat{\Pi}(\bm{x})\right) ,
\end{align}
which satisfy $\left[ \hat{q}_{A},\hat{p}_{A}\right] =i$, and the weighting
functions of $A$, $x_{A}(\bm{x}),y_{A}(\bm{x}),z_{A}(\bm{x})$ and $w_{A}(\bm{x})$, are real functions localized in a spatial region. By varying $f\left( q,p\right) =\hat{q}_{A}^{n}\hat{p}_{A}^{m}+\hat{p}_{A}^{m}\hat{q}_{A}^{n}$ with non-negative integers $m$ and $n$, the detector is capable of measuring multipoint correlation
functions $\langle \Psi |\left( \hat{q}_{A}^{n}\hat{p}_{A}^{m}+\hat{p}_{A}^{m}\hat{q}_{A}^{n}\right) |\Psi \rangle $ of $\hat{q}_{A}$ and $\hat{p}_{A}$. The entire measurement result of the correlation functions for all $n$
and $m$ can be summarized as a generating function $\chi _{A}\left(
x_{A},v_{A}\right) =\langle \Psi |e^{i\left( v_{A}\hat{q}_{A}-x_{A}\hat{p}_{A}\right) }|\Psi \rangle $ of the correlation functions. The components of
the reduced state $\hat{\rho}_{A}$ of $A$ in the position basis can be
determined by the measured function $\chi _{A}\left( x_{A},v_{A}\right) $ as
follows:

\begin{equation*}
\langle \bar{x}_{A}|\hat{\rho}_{A}|x_{A}\rangle =\frac{1}{(2\pi )^{2}}\int
\chi _{A}\left( x_{A}-\bar{x}_{A},v_{A}\right) e^{-\frac{i}{2}v_{A}\left( 
\bar{x}_{A}+x_{A}\right) }dv_{A}.
\end{equation*}
The proof that $\hat{\rho}_{A}$ actually becomes non-negative Hermitian
operator satisfying normalization condition, $\mathrm{Tr}\left[ \hat{\rho}_{A}\right] =1$ is given in Appendix \ref{append_tilderho}. Because the mode $A$ is coupled to
other modes in general, $\hat{\rho}_{A}$ usually becomes a mixed state.
Since the entire field is in a pure state, there exists a purification partner mode $B$ of $A$, and the $AB$ system is in a pure
entangled state $\ket{\psi}_{AB}$. Taking the partial trace of $B$, the
reduced state $\hat{\rho}_{A}$ is reproduced as $\hat{\rho}_{A}=\mathrm{Tr}_{B}\left[ \ket{\psi}_{AB}\bra{\psi}_{AB}\right] $. Now,
we propose a generalized definition of partner mode. The partner mode $B$ of 
$A$ is characterized by a set of operators $(\hat{q}_{B},\hat{p}_{B})$
satisfying the following conditions (i), (ii) and (iii):

\begin{enumerate}[(i)]
\item \textbf{Commutation relation}: $[\hat{q}_B,\hat{p}_B]=i$.

\item \textbf{Locality}: $[\hat{q}_A,\hat{q}_B]=0$, $[\hat{q}_A,\hat{p}_B]=0$, $[\hat{p}_A,\hat{q}_B]=0$, and $[\hat{p}_A,\hat{p}_B]=0$.

\item \textbf{Purification condition}: The correlation space state $\hat{\tilde{\rho}}
_{AB}$ whose components in the position basis are given by 
\begin{align}
& \Braket{\bar{x}_A,\bar{x}_B|\hat{\tilde{\rho}}_{AB}|x_A,x_B}  \notag \\
& \equiv \frac{1}{(2\pi )^{2}}\int dv_{A}dv_{B}\chi \left( x_{A}-\bar{x}_{A},v_{A},x_{B}-\bar{x}_{B},v_{B}\right) e^{-\frac{i}{2}\left( v_{A}\left( 
\bar{x}_{A}+x_{A}\right) +v_{B}\left( \bar{x}_{B}+x_{B}\right) \right) }
\label{eq_Wigner}
\end{align}
is pure. Here, we have used the Wigner characteristic function defined by 
\begin{equation*}
\chi (x_{A},v_{A},x_{B},v_{B})\equiv \langle \Psi |e^{i\left( v_{A}\hat{q}_{A}-x_{A}\hat{p}_{A}\right) }e^{i\left( v_{B}\hat{q}_{B}-x_{B}\hat{p}_{B}\right) }|\Psi \rangle 
\end{equation*}
for the pure state $|\Psi \rangle $ of system.  
\end{enumerate}

\bigskip 

Though $(\hat{q}_{A},\hat{p}_{A})$ are assumed to be a linear combination of 
$\hat{\varphi}(x)$ and $\hat{\Pi}(x)$, $(\hat{q}_{B},\hat{p}_{B})$ are not.
The partner operators $(\hat{q}_{B},\hat{p}_{B})$ can include non-linear
terms like $\hat{\varphi}(x)^{n}$ and $\hat{\Pi}(x)^{m}$ in general.\ The
condition (ii) ensures the locality necessary to introduce the notion of
entanglement, while (iii) gives the condition that the partner $(\hat{q}_{B}, \hat{p}_{B})$ purifies $(\hat{q}_{A},\hat{p}_{A})$. As well as $\hat{\rho}_{A}$, $\hat{\tilde{\rho}}_{AB}$ is a quantum state, i.e., a unit trace
positive-semidefinite Hermitian operator. In order to introduce the concept
of entanglement, $\hat{\tilde{\rho}}_{AB}$ is determined by the correlation
functions of local operators of $A$ and $B$. The Wigner characteristic
function $\chi (x_{A},v_{A},x_{B},v_{B})$ actually satisfies this postulate
and yields all the correlation functions. Thus, our definition works well.
Another necessary condition for $\hat{\tilde{\rho}}_{AB}$ is the following:
for a state $\hat{\rho}$ of two harmonic oscillator system, $\hat{\tilde{\rho}}_{AB}=\hat{\rho}$ must hold. Eq. (\ref{eq_Wigner}) actually obeys this
condition which can be confirmed by using the Fourier transformation and its
inverse transformation simultaneously. If we have a partner candidate $B$
with $(\hat{q}_{B},\hat{p}_{B})$, experimental measurements of the correlation functions
of $(\hat{q}_{A},\hat{p}_{A},\hat{q}_{B},\hat{p}_{B})$ allow us to corroborate the partner of $A$ in principle.

Since for a Gaussian state, the Wigner characteristic function is fully
characterized by a $4\times 4$ matrix called the covariance matrix, the
condition (iii) gets simplified as explained in the next section.

By using the result on a pair of partners $A$ and $B$ for the Gaussian
vacuum state $\ket{0}$ of a field, nontrivial examples of partners for
non-Gaussian states can be easily constructed. Let us consider a general
unitary operation $\hat{U}$ generated by a non-linear interaction
Hamiltonian consist of $\hat{\phi}$ and $\hat{\Pi}$. The post-operated state 
$\ket{\Psi}=\hat{U}\ket{0}$ is non-Gaussian. In the state, we have partners
which are defined as $\left( \hat{q}_{A}^{\prime },\hat{p}_{A}^{\prime
}\right) =\left( \hat{U}\hat{q}_{A}\hat{U}^{\dag },\hat{U}\hat{p}_{A}\hat{U}^{\dag }\right) $ and $\left( \hat{q}_{B}^{\prime },\hat{p}_{B}^{\prime
}\right) =\left( \hat{U}\hat{q}_{B}\hat{U}^{\dag },\hat{U}\hat{p}_{B}\hat{U}^{\dag }\right) $. The characteristic function becomes the same as that of
the corresponding Gaussian partners: 
\begin{equation*}
\mathrm{Tr}\left( \Ket{\Psi}\Bra{\Psi}e^{i\left( v_{A}\hat{q}_{A}^{\prime
}-x_{A}\hat{p}_{A}^{\prime }\right) }e^{i\left( v_{B}\hat{q}_{B}^{\prime
}-x_{B}\hat{p}_{B}^{\prime }\right) }\right) =\mathrm{Tr}\left( \ket{0}\bra{0}e^{i\left( v_{A}\hat{q}_{A}-x_{A}\hat{p}_{A}\right) }e^{i\left( v_{B}\hat{q}_{B}-x_{B}\hat{p}_{B}\right) }\right) .
\end{equation*}
Thus, $\left( \hat{q}_{A}^{\prime },\hat{p}_{A}^{\prime }\right) $ and $\left( \hat{q}_{B}^{\prime },\hat{p}_{B}^{\prime }\right) $ provide partners
for a quantum field in a pure non-Gaussian state $\Ket{\Psi}$. From the
viewpoint of pure mathematics, the example is merely a unitary-equivalent
one to partners in Gaussian states. However it should be stressed that the above example is nontrivial in a physical sense. The above particle modes in the non-Gaussian state are physically detected by realistic particle detectors which fix what operators can be observed.

Beyond the above example, a natural question arises: If we fix an arbitrary
mode $A$ of a field in a general state, does its partner always exist?
Interestingly the answer is "yes" when we consider $N$ coupled harmonic
oscillators as a $1+1$ dimensional discretized scalar quantum field in a
general pure state $\ket{\Psi}_{1,\cdots,N}$. Let us define a
particle mode $A$ as a linear combination: 
\begin{equation}
\hat{q}_{A}\equiv \sum_{n=1}^{N}\left( x_{A}(n)\hat{q}_{n}+y_{A}(n)\hat{p}_{n}\right) ,\quad \hat{p}_{A}\equiv \sum_{n=1}^{N}\left( z_{A}(n)\hat{q}_{n}+w_{A}(n)\hat{p}_{n}\right) , \label{300}
\end{equation}
where $(\hat{q}_n,\hat{p}_n)$ denote the canonical operators for $n$th harmonic oscillator. Imposing the condition $[\hat{q}_{A},\hat{p}_{A}]=i$, we have a constraint on the coefficients: 
\begin{equation*}
\sum_{n=1}^{N}\left( x_{A}(n)w_{A}(n)-z_{A}(n)y_{A}(n)\right) =1.
\end{equation*}
The Stone-von Neumann theorem \cite{Hall} guarantees that there
exists an unitary operator $\hat{V}_{N}$ such that $\hat{V}_{N}\hat{q}_{A}\hat{V}_{N}^{\dag }=\hat{q}_{1}$ and $\hat{V}_{N}\hat{p}_{A}\hat{V}_{N}^{\dag }=\hat{p}_{1}$. The transformed state is given by $\ket{\Psi'}_{1,\cdots, N}\equiv \hat{V}\ket{\Psi}_{1,\cdots,N}$ and remains pure. Let us consider the
Schmidt decomposition of $|\Psi ^{\prime }\rangle $ as 
\begin{equation*}
|\Psi ^{\prime }\rangle_{1,\cdots, N} =\sum_{n=0}^{\infty }\sqrt{p_{n}}|a_{n}\rangle_1 |\psi
_{n}\rangle_{2,\cdots N} ,
\end{equation*}
where $\{p_{n}\}_{n=0}^\infty$ is a probability distribution. Here we assume that the reduced state $\hat{\rho}_{1}$ of the first mode, which is defined as 
\begin{equation*}
\hat{\rho}_{1}=\mathrm{Tr}_{2\cdots N}\left[ |\Psi ^{\prime }\rangle
\langle \Psi ^{\prime }|\right] ,
\end{equation*}
has a spectral decomposition in terms of a discrete basis $\left\{
|a_{n}\rangle :n=0,1,2,\cdots \right\} $ of the sub-Hilbert space as 
\begin{equation*}
\hat{\rho}_{1}=\sum_{n=0}^{\infty }p_{n}|a_{n}\rangle _1\langle a_{n}|_1.
\end{equation*}
This may be not an essential constraint, and if the continuum spectrum
emerges, a small modification and generalization of this argument is
expected to yield the same conclusions. To obtain the partner mode, let us consider the following creation and annihilation operators:
\begin{align*}
 \hat{b}^\dag \equiv \sum_{n=0}^{\infty}\sqrt{n+1}\sum_{i=1}^{\infty}\ket{\psi_{n+1}^{(i)}}_{2,\cdots,N}\bra{\psi_n^{(i)}}_{2,\cdots,N}, \quad \hat{b}\equiv \sum_{n=0}^\infty \sqrt{n+1}\sum_{i=1}^{\infty}\ket{\psi_n^{(i)}}_{2,\cdots,N}\bra{\psi_{n+1}^{(i)}}_{2,\cdots,N},
\end{align*}
where we have introduced an orthonormal basis $\{\ket{\psi_n^{(i)}}\}$, satisfying $\ket{\psi_n^{(1)}}=\ket{\psi_n}$ for all $n$, and $\Braket{\psi_n^{(i)}|\psi_m^{(j)}}=\delta_{nm}\delta_{ij}$.
These operators satisfy $[\hat{b},\hat{b}^\dag]=1$. Then, the conditions (i), (ii) and (iii) are satisfied for $(\hat{q}_B,\hat{p}_B)$ defined by
\begin{align}
 \hat{q}_B\equiv \hat{V}_N^\dag \left(\hat{I}\otimes \frac{1}{\sqrt{2}}\left(\hat{b}+\hat{b}^\dag\right)\right) \hat{V}_N, \quad  \hat{p}_B\equiv \hat{V}_N^\dag \left(\hat{I}\otimes \frac{1}{\sqrt{2}i}\left(\hat{b}-\hat{b}^\dag\right)\right) \hat{V}_N.
\end{align}
Regarding this construction of the partner, the following three points should be noted. First, such a partner is non-unique if $\hat{\rho}_1$ is not full rank. For example, if $p_0=0$, $\ket{\psi_0}_{2,\cdots,N}$ can be an arbitrary normalized vector orthogonal to $\ket{\psi_n}_{2,\cdots,N}$ ($n=1,2,\cdots$). Furthermore, even when $\hat{\rho}_1$ is full rank, $\{\ket{\psi_n^{(i)}}\}_{i=2}^{N-1}$ can be an arbitrary set of orthonormal vectors as long as $\{\ket{\psi_n^{(i)}}\}_{i=1}^{N-1}$ forms an orthonormal basis. Second, $(\hat{q}_B,\hat{p}_B)$ may not be linear combinations of $\hat{q}_n$ and $\hat{p}_n$. Third, the continuum limit to reproduce the original field remains subtle and requires further delicate analysis. Nevertheless, surprisingly, it is shown that for Gaussian states, there exists the unique partner whose canonical operators are given by linear combination of $\hat{q}_n$ and $\hat{p}_n$. A closed formula to obtain the partner mode is presented in the following section. In this case, it is possible to take the continuum limit, i.e., we have the unique partner for the free scalar field in Gaussian states.

\section{Partner mode in the Gaussian vacuum States}\label{sec_partner_vac} 
In this section, we derive the partner formula for a
Gaussian vacuum state of a free scalar field. The extension of the formula
for an excited Gaussian state is given in the following section. We first derive the partner formula for a
discretized scalar quantum field theory in a flat $(1+1)$-dimensional
spacetime. Let us impose a periodic boundary condition on the field: 
\begin{equation*}
\hat{\phi}(t,x+L)=\hat{\phi}(t,x),
\end{equation*}
where $L$ denotes the entire space length. The free Hamiltonian of the
system is given by 
\begin{equation*}
\hat{H}=\frac{1}{2}\int_{-L/2}^{L/2}dx:\hat{\Pi}(x)^{2}:+\frac{1}{2}\int_{-L/2}^{L/2}dx:\left( \partial _{x}\hat{\phi}(x)\right) ^{2}:+\frac{m^{2}}{2}\int_{-L/2}^{L/2}dx:\hat{\phi}(x)^{2}:,
\end{equation*}
where $:\hat{\mathcal{O}}:$ is the normal ordering of a linear operator $\hat{\mathcal{O}}$, and $\hat{\Pi}(x)$ is the canonical momentum of the
field $\hat{\phi}(x)$ satisfying 
\begin{equation*}
\left[ \hat{\phi}(x),\hat{\Pi}(x^{\prime })\right] =i\delta (x-x^{\prime }).
\end{equation*}

In order to obtain the partner formula, consider a corresponding discretized
model with lattice spacing $\epsilon$. The field operator $\hat{\phi}(x)$
and its conjugate momentum $\hat{\Pi}(x)$ correspond to 
\begin{align}
\hat{\phi}(x)\to\frac{\hat{q}_n}{\sqrt{m\epsilon}},\quad \hat{\Pi}(x)\to 
\sqrt{\frac{m}{\epsilon}}\hat{p}_n .
\end{align}

Introducing new variables $N\equiv L/\epsilon $ and $\eta \equiv
1/(m\epsilon )^{2}$ reproduces the discretized Hamiltonian of the coupled
harmonic oscillators:
\begin{equation}
\hat{H}=\frac{1}{2}\sum_{n=1}^{N}:\hat{p}_{n}^{2}:+\left( \frac{1}{2}+\eta
\right) \sum_{n=1}^{N}:\hat{q}_{n}^{2}:-\eta \sum_{n=1}^{N}:\hat{q}_{n+1}\hat{q}_{n}: \label{301}
\end{equation}
where $:\hat{O}:$ means normal ordered operator of $\hat{O}$ with respect to
creation and annihilation operators, $\hat{q}_{m}$ and $\hat{p}_{n}$ satisfy
the canonical commutation relations $\left[ \hat{q}_{m},\hat{p}_{n}\right]
=i\delta _{mn}$. The
Hamiltonian generates the evolution with respect to a new time coordinate $\tau \equiv mt$ . By using the mode functions 
\begin{equation*}
u_{k}(n)\equiv \frac{1}{\sqrt{N}}\exp {\left( 2\pi ik\frac{n}{N}\right) },
\end{equation*}
the canonical operators are expanded as 
\begin{align}
\hat{q}_{n}& =\sum_{k=0}^{N-1}\frac{1}{\sqrt{2\omega _{k}}}\left( \hat{a}_{k}u_{k}(n)+\hat{a}_{k}^{\dag }u_{k}(n)^{\ast }\right) ,\quad \hat{p}_{n} =\frac{1}{i}\sum_{k=0}^{N-1}\sqrt{\frac{\omega _{k}}{2}}\left( 
\hat{a}_{k}u_{k}(n)-\hat{a}_{k}^{\dag }u_{k}(n)^{\ast }\right) \label{eq_qp},
\end{align}
where the dispersion relationship is given by 
\begin{equation*}
\omega _{k}^{2}=1+2\eta \left( 1-\cos \left( \frac{2\pi k}{N}\right) \right)
.
\end{equation*}
Since the canonical commutation relation $[\hat{q}_{n},\hat{p}_{m}]=i\delta
_{mn}$ yields $[\hat{a}_{k},\hat{a}_{k^{\prime }}^{\dag }]=\delta
_{kk^{\prime }}$, $\hat{a}_{k}^{\dag }$ and $\hat{a}_{k}$ are creation and
annihilation operator for a mode $k$. The vacuum state $\ket{0}$ is defined
as a unit vector annihilated by $\hat{a}_{k}$ for all $k=0,1,\cdots N-1$.
Hereafter, $\Braket{\hat{\mathcal{O}}}\equiv \Braket{0|\hat{\mathcal{O}}|0}$
for a linear operator $\hat{\mathcal{O}}$. Let us consider a set of
canonical variables $(\hat{q}_{A},\hat{p}_{A})$ in the previous section.
For the derivation of the partner formula, we will use the covariance matrix.
For a review of its properties, see Appendix \ref{covariance}. The covariance
matrix associated to the canonical variables $(q_{A},p_{A})$ is given by 
\begin{equation*}
m_{A}=
\begin{pmatrix}
\Braket{\hat{q}_A^2} & \mathrm{Re}\left( \Braket{\hat{q}_A\hat{p}_A}\right) 
\\ 
\mathrm{Re}\left( \Braket{\hat{p}_A\hat{q}_A}\right)  & \Braket{\hat{p}_A^2}
\end{pmatrix}
.
\end{equation*}

Through a local symplectic transformation 
\begin{align}
\begin{pmatrix}
\hat{Q}_A \\ 
\hat{P}_A
\end{pmatrix}
=S_A 
\begin{pmatrix}
\hat{q}_A \\ 
\hat{p}_A
\end{pmatrix}
= 
\begin{pmatrix}
\cos\theta_A^{\prime } & \sin\theta_A^{\prime } \\ 
-\sin\theta_A^{\prime } & \cos\theta_A^{\prime }
\end{pmatrix}
\begin{pmatrix}
e^{\sigma_A} & 0 \\ 
0 & e^{-\sigma_A}
\end{pmatrix}
\begin{pmatrix}
\cos\theta_A & \sin\theta_A \\ 
-\sin\theta_A & \cos\theta_A
\end{pmatrix}
\begin{pmatrix}
\hat{q}_A \\ 
\hat{p}_A
\end{pmatrix}
\end{align}
with $\theta_A^{\prime },\sigma_A,\theta_A\in\mathbb{R}$, it is possible to
bring the covariance matrix $M_A$ for new canonical variables $(\hat{Q}_A,\hat{P}_A)$ to the following standard form: 
\begin{align}
M_A= 
\begin{pmatrix}
\Braket{\hat{Q}_A^2} & \mathrm{Re}\left(\Braket{\hat{Q}_A\hat{P}_A}\right)
\\ 
\mathrm{Re}\left(\Braket{\hat{P}_A\hat{Q}_A}\right) & \Braket{\hat{P}_A^2}
\end{pmatrix}
= \frac{\sqrt{1+g^2}}{2} 
\begin{pmatrix}
1 & 0 \\ 
0 & 1
\end{pmatrix}
,  \label{sym_localA}
\end{align}
where $g$ is a non-negative parameter. As is easily seen, we can take $\theta_A^{\prime }=0$ without loss of generality. Solving Eq.~(\ref{sym_localA}), $\sigma_A$ and $\theta_A$ are fixed. Note that this $g$ is
uniquely determined by the elements of $m_A$ since 
\begin{align}
\frac{1}{4}\left(1+g^2\right)=\det{M_A}=\det{m_A}=\Braket{\hat{q}_A^2}\Braket{\hat{p}_A^2}-\frac{1}{4}\left(\Braket{\hat{q}_A\hat{p}_A}+\Braket{\hat{p}_A\hat{q}_A}\right)^2,  \label{eq_g_det}
\end{align}
where we have used $\det{S_A}\det{S_A\sps{T}}=1$. When $g=0$, the mode $A$ is in a pure
state. If $g\neq0$, the mode $A$ is in a mixed state, meaning that there is
a purification partner mode $B$. Hereafter, we assume $g\neq 0$, i.e., the
mode $A$ is in a mixed state. In the following, we will construct a set of
canonical variables $(\hat{Q}_B,\hat{P}_B)$ that represents the purification
partner of mode $A$ such that the composite system $AB$ is in a pure state.
This purification partner $B$ of $A$ is characterized by a set of canonical
variables 
\begin{align}
\hat{Q}_B=\sum_{n=1}^{N}\left(X_B(n)\hat{q}_n+Y_B(n) \hat{p}_n\right),\quad 
\hat{P}_B=\sum_{n=1}^{N}\left(Z_B(n)\hat{q}_n+W_B(n)\hat{p}_n\right)
\end{align}
which must satisfy the following:

\begin{enumerate}[(i)]

\item \textbf{Commutation relation:} $\left[\hat{Q}_B,\hat{P}_B\right]=i$

\item \textbf{Locality:} $\left[\hat{Q}_A,\hat{Q}_B\right]=0$, $\left[\hat{Q}
_A,\hat{P}_B\right]=0$, $\left[\hat{P}_A,\hat{Q}_B\right]=0$, and $\left[\hat{P}_A,\hat{P}_B\right]=0$

\item \textbf{Purification condition:} the covariance matrix takes the
following form: 
\begin{align}
M_{AB}&\equiv 
\begin{pmatrix}
\Braket{\hat{Q}_A^2} & \mathrm{Re}\left(\Braket{\hat{Q}_A\hat{P}_A}\right) & 
\Braket{\hat{Q}_A\hat{Q}_B} & \Braket{\hat{Q}_A\hat{P}_B} \\ 
\mathrm{Re}\left(\Braket{\hat{P}_A\hat{Q}_A}\right) & \Braket{\hat{P}_A^2} &
\Braket{\hat{P}_A\hat{Q}_B} & \Braket{\hat{P}_A\hat{P}_B} \\ 
\Braket{\hat{Q}_B\hat{Q}_A} & \Braket{\hat{Q}_B\hat{P}_A} & 
\Braket{\hat{Q}_B^2} & \mathrm{Re}\left(\Braket{\hat{Q}_B\hat{P}_B}\right)
\\ 
\Braket{\hat{P}_B\hat{Q}_A} & \Braket{\hat{P}_B\hat{P}_A} & \mathrm{Re}\left(\Braket{\hat{P}_B\hat{Q}_B}\right) & \Braket{\hat{P}_B^2}
\end{pmatrix}
\notag \\
&= 
\begin{pmatrix}
\frac{1}{2}\sqrt{1+g^2} & 0 & \frac{g}{2} & 0 \\ 
0 & \frac{1}{2}\sqrt{1+g^2} & 0 & -\frac{g}{2} \\ 
\frac{g}{2} & 0 & \frac{1}{2}\sqrt{1+g^2} & 0 \\ 
0 & -\frac{g}{2} & 0 & \frac{1}{2}\sqrt{1+g^2}
\end{pmatrix}
,  \label{cov_mat_AB}
\end{align}
such that the state of the composite system $AB$ is in a pure state.
\end{enumerate}

The condition (iii) in Section \ref{sec_definition} is now simplified to a
condition on the covariance matrix for a two-mode Gaussian state. Eq.~(\ref{cov_mat_AB}) is what is called the standard form of the covariance matrix
for a pure Gaussian state \cite{S, DGCZ}. More details about the covariance
matrix can be found in Appendix \ref{covariance}. As we will see, the
purification condition plays a crucial role to obtain the partner formula.

Since the operators $\left(\hat{Q}_A,\hat{P}_A,\hat{Q}_B,\hat{P}_B\right)$
are constructed as a linear combination of $\{(\hat{q}_n,\hat{p}_n)\}_{n=1}^N $, any expectation value of a product of operators $\left(\hat{Q}_A,\hat{P}_A,\hat{Q}_B,\hat{P}_B\right)$ for a Gaussian state is
calculated by using the Wick's theorem. Thus, the covariance matrix $M_{AB}$
characterizes arbitrary observable on the two-mode system $AB$, meaning that
it gives a reduced state in the correlation space.

On the other hand, the locality conditions imply that no operation $\hat{U}_{B}\left( \hat{Q}_{B},\hat{P}_{B}\right) $ on mode $B$ generated by $\left( 
\hat{Q}_{B},\hat{P}_{B}\right)$ affects the reduced state of mode $A$, and
vice-versa. Therefore, in the correlation space spanned by $\left( \hat{Q}_{A},\hat{P}_{A},\hat{Q}_{B},\hat{P}_{B}\right) $ \cite{cc1} \cite{cc2}, $A$
and $B$ are locally independent. Since locality of $A$ and $B$ can be
introduced, quantum entanglement among $A$ and $B$ is well defined. The
entanglement entropy $S_{\mathrm{EE}} $ between mode $A$ and its partner $B$
depends on the positive parameter $g$ as follows \cite{HSH}: 
\begin{align}
S_{\mathrm{EE}} =\sqrt{1+g^2}\ln{\left(\frac{1}{g}\left(\sqrt{1+g^2}+1\right)\right)}+\ln{\left(\frac{g}{2}\right)}.  \label{eq_EE}
\end{align}
The purification condition on equation \eqref{cov_mat_AB} can be summarized
as follows: 
\begin{align}
& \Braket{\hat{Q}_A\hat{P}_B}=\Braket{\hat{P}_A\hat{Q}_B}=0,
\label{partner_cond_1} \\
&\Braket{\hat{Q}_A\hat{Q}_B}=-\Braket{\hat{P}_A\hat{P}_B}= \frac{g}{2}, \\
&\mathrm{Re}\left(\Braket{\hat{Q}_B\hat{P}_B}\right)=0, \\
&\Braket{\hat{Q}_B^2}=\Braket{\hat{P}_B^2}=\frac{\sqrt{1+g^2}}{2} .
\end{align}
In addition, the commutation relation $[\hat{Q}_B,\hat{P}_B]=i$ gives us: 
\begin{align}
\Braket{\hat{Q}_B\hat{P}_B-\hat{P}_B\hat{Q}_B}=i.  \label{partner_cond_2}
\end{align}

To obtain the solution of the above equations, let us expand $(\hat{Q}_A,\hat{P}_A,\hat{Q}_B,\hat{P}_B)$ in terms of $\hat{a}_k$ and $\hat{a}_k^\dag$
as follows: 
\begin{align}
\hat{Q}_A&=\left(\frac{\sqrt{1+g^2}}{2}\right)^{1/2}\sum_{k=0}^{N-1}\left(Q_A(k)^*\hat{a}_k+Q_A(k)\hat{a}_k^\dag\right), \\
\hat{P}_A&=\left(\frac{\sqrt{1+g^2}}{2}\right)^{1/2}\sum_{k=0}^{N-1}\left(P_A(k)^*\hat{a}_k+P_A(k)\hat{a}_k^\dag\right), \\
\hat{Q}_B&=\left(\frac{\sqrt{1+g^2}}{2}\right)^{1/2}\sum_{k=0}^{N-1}\left(Q_B(k)^*\hat{a}_k+Q_B(k)\hat{a}_k^\dag\right), \\
\hat{P}_B&=\left(\frac{\sqrt{1+g^2}}{2}\right)^{1/2}\sum_{k=0}^{N-1}\left(P_B(k)^*\hat{a}_k+P_B(k)\hat{a}_k^\dag\right),
\end{align}
where we have factored out $\left(\frac{\sqrt{1+g^2}}{2}\right)^{1/2}$ for
future convenience.

For Hermite operators $\hat{O}_1$ and $\hat{O}_2$ defined as linear
combinations of $a_k$ and $a_k^\dag$ such as 
\begin{align}
\hat{O}_i&=\left(\frac{\sqrt{1+g^2}}{2}\right)^{1/2}\sum_{k=0}^{N-1}\left(O_i(k)^*\hat{a}_k+O_i(k)\hat{a}_k^\dag\right),
\end{align}
we get 
\begin{align}
\Braket{\hat{O}_1\hat{O}_2}=\frac{\sqrt{1+g^2}}{2}\Braket{O_1,O_2},
\end{align}
where we have defined the standard inner product in $\mathbb{C}^N$: 
\begin{align}
\Braket{O_1,O_2}\equiv \sum_{k=0}^{N-1}O_1(k)^*O_2(k).
\end{align}

Eqs.~(\ref{partner_cond_1})-(\ref{partner_cond_2}) are expressed as the
followings: 
\begin{align}
\begin{pmatrix}
\Braket{Q_A,Q_A} & \Braket{Q_A,P_A} & \Braket{Q_A,Q_B} & \Braket{Q_A,P_B} \\ 
\Braket{P_A,Q_A} & \Braket{P_A,P_A} & \Braket{P_A,Q_B} & \Braket{P_A,P_B} \\ 
\Braket{Q_B,Q_A} & \Braket{Q_B,P_A} & \Braket{Q_B,Q_B} & \Braket{Q_B,P_B} \\ 
\Braket{P_B,Q_A} & \Braket{P_B,P_A} & \Braket{P_B,Q_B} & \Braket{P_B,P_B}
\end{pmatrix}
= 
\begin{pmatrix}
1 & \frac{i}{\sqrt{1+g^2}} & \frac{g}{\sqrt{1+g^2}} & 0 \\ 
-\frac{i}{\sqrt{1+g^2}} & 1 & 0 & -\frac{g}{\sqrt{1+g^2}} \\ 
\frac{g}{\sqrt{1+g^2}} & 0 & 1 & \frac{i}{\sqrt{1+g^2}} \\ 
0 & -\frac{g}{\sqrt{1+g^2}} & -\frac{i}{\sqrt{1+g^2}} & 1
\end{pmatrix}
.
\end{align}

Since $Q_A$ and $P_B$ are orthonormal, $\left|\Braket{Q_A,Q_B}\right|^2+\left|\Braket{P_B,Q_B}\right|^2=1 $ and $\left|\braket{Q_B,Q_B}\right|^2=1$ imply that 
\begin{align}
Q_B(k)&=\Braket{Q_A,Q_B}Q_A(k)+\Braket{P_B,Q_B}P_B(k)  \notag \\
&= \frac{g}{\sqrt{1+g^2}}Q_A(k)-\frac{i}{\sqrt{1+g^2}}P_B(k).
\label{partner_deri_QB}
\end{align}
Similarly, 
\begin{align}
P_B(k)&=-\frac{g}{\sqrt{1+g^2}}P_A(k)+\frac{i}{\sqrt{1+g^2}}Q_B(k).
\label{partner_deri_PB}
\end{align}
Combining Eqs.~(\ref{partner_deri_QB}) and (\ref{partner_deri_PB}), we
finally get the unique solution: 
\begin{equation}
Q_B(k)=\frac{\sqrt{1+g^2}}{g}Q_A(k)+\frac{i}{g}P_A(k),\quad P_B(k)=-\frac{\sqrt{1+g^2}}{g}P_A(k) +\frac{i}{g}Q_A(k).
\end{equation}
It should be noted that the commutativity condition among $(\hat{Q}_A,\hat{P}_A)$ and $(\hat{Q}_B,\hat{P}_B)$ automatically satisfied since 
\begin{align}
\left[\hat{\mathcal{O}}_1,\hat{\mathcal{O}}_2\right]=\frac{\sqrt{1+g^2}}{2}\left(\Braket{\mathcal{O}_1,\mathcal{O}_2}-\Braket{\mathcal{O}_2,\mathcal{O}_1}\right).
\end{align}

Therefore, the partner mode is written as 
\begin{align}
\hat{Q}_{B}& =\frac{\sqrt{1+g^{2}}}{g}\hat{Q}_{A}-\frac{i}{g}\left( \frac{\sqrt{1+g^{2}}}{2}\right) ^{1/2}\sum_{k=0}^{N-1}\left( P_{A}(k)^{\ast }\hat{a}_{k}-P_{A}(k)a_{k}^{\dag }\right) , \\
\hat{P}_{B}& =-\frac{\sqrt{1+g^{2}}}{g}\hat{P}_{A}-\frac{i}{g}\left( \frac{\sqrt{1+g^{2}}}{2}\right) ^{1/2}\sum_{k=0}^{N-1}\left( Q_{A}(k)^{\ast }\hat{a}_{k}-Q_{A}(k)\hat{a}_{k}^{\dag }\right) .
\end{align}
By re-writing the last equation in term of the weighting functions: 
\begin{eqnarray}
X_{A}(n) &=&\sum_{k=0}^{N-1}\sqrt{\frac{\omega _{k}}{2}}\left[ Q_{A}^{\ast
}(k)u_{k}^{\ast }(n)+Q_{A}(k)u_{k}(n)\right] \ , \\
Y_{A}(n) &=&\sum_{k=0}^{N-1}\imath \frac{1}{\sqrt{2\omega _{k}}}\left[
Q_{A}^{\ast }(k)u_{k}^{\ast }(n)-Q_{A}(k)u_{k}(n)\right] \ , \\
Z_{A}(n) &=&\sum_{k=0}^{N-1}\sqrt{\frac{\omega _{k}}{2}}\left[ P_{A}^{\ast
}(k)u_{k}^{\ast }(n)+P_{A}(k)u_{k}(n)\right] \ , \\
W_{A}(n) &=&\sum_{k=0}^{N-1}\imath \frac{1}{\sqrt{2\omega _{k}}}\left[
P_{A}^{\ast }(k)u_{k}^{\ast }(n)-P_{A}(k)u_{k}(n)\right] \ ,
\end{eqnarray}
and similarly for the partner $B$ weighting functions, the partner $B$ can
be written in terms of the weighting functions of mode $A$ as follows: 
\begin{align}
\hat{Q}_{B}& =\left( \frac{\sqrt{1+g^{2}}}{2}\right)
^{1/2}\sum_{n=1}^{N}\left( X_{B}(n)\hat{q}_{n}+Y_{B}(n)\hat{p}_{n}\right) ,
\\
\hat{P}_{B}& =\left( \frac{\sqrt{1+g^{2}}}{2}\right)
^{1/2}\sum_{n=1}^{N}\left( Z_{B}(n)\hat{q}_{n}+W_{B}(n)\hat{p}_{n}\right) ,
\end{align}
where 
\begin{align}
X_{B}(n)& \equiv \frac{\sqrt{1+g^{2}}}{g}X_{A}(n)-\frac{2}{g}\sum_{n^{\prime
}=1}^{N}\Delta _{p}(n-n^{\prime })W_{A}(n^{\prime }),
\label{eq_partner_in_vacuum_dis_1} \\
Y_{B}(n)& \equiv \frac{\sqrt{1+g^{2}}}{g}Y_{A}(n)+\frac{2}{g}\sum_{n^{\prime
}=1}^{N}\Delta _{q}(n-n^{\prime })Z_{A}(n^{\prime }), \\
Z_{B}(n)& \equiv -\frac{\sqrt{1+g^{2}}}{g}Z_{A}(n)-\frac{2}{g}\sum_{n^{\prime }=1}^{N}\Delta _{p}(n-n^{\prime })Y_{A}(n^{\prime }), \\
W_{B}(n)& \equiv -\frac{\sqrt{1+g^{2}}}{g}W_{A}(n)+\frac{2}{g}\sum_{n^{\prime }=1}^{N}\Delta _{q}(n-n^{\prime })X_{A}(n^{\prime }),
\label{eq_partner_in_vacuum_dis_2}
\end{align}
with 
\begin{align}
\Delta _{q}(n-n^{\prime })& \equiv \Braket{\hat{q}_n \hat{q}_{n'}}=\frac{1}{N}\sum_{k=0}^{N-1}\frac{1}{2\omega _{k}}\exp {\left( 2\pi ik\frac{n-n^{\prime
}}{N}\right) }, \\
\Delta _{p}(n-n^{\prime })& \equiv \Braket{\hat{p}_n \hat{p}_{n'}}=\frac{1}{N}\sum_{k=0}^{N-1}\frac{\omega _{k}}{2}\exp {\left( 2\pi ik\frac{n-n^{\prime }}{N}\right) }.
\end{align}
This is the partner formula for the vacuum of the free lattice scalar field
theory. Before taking the continuum limit, let us analyze our results. From
our partner formula, two different kinds of partner can be defined: the
spatially separated partner (SSP) and spatially overlapped partner (SOP) as
follows:

\begin{definition*}
If the weighting functions of mode $B$: $\{X_{B}(n),Y_{B}(n),Z_{B}(n),W_{B}(n)\}_{n}$ have any spatial overlap with $\{X_{A}(n),Y_{A}(n),Z_{A}(n),W_{A}(n)\}_{n}$, we call the modes $A$ and $B$
spatially overlapped partners (SOP). If not, we call them spatially
separated partners (SSP).
\end{definition*}

This definition is straightforwardly extended for an arbitrary Gaussian
state in the scalar field theory. In \cite{BR1,BR2}, SSPs have been
constructed for a special case to investigate the spatial structure of
entanglement in the vacuum state. By using our partner formula, it is
possible to investigate not only SSPs but also SOPs. Thus, it provides a new
way to extract and make use of information stored in a quantum field.
Furthermore, since one can identify the partner mode $B$ for arbitrary mode $A$, it can be used to introduce a tensor product structure in the entire
Hilbert space even when there is not a natural tensor product structure in
advance.

So far, we have obtained the partner formula in a $(1+1)$-dimensional
lattice free field theory. The extension of the results into a $(d+1)$-dimensional spacetime is obtained in a straightforward way. First, let us
extend our results to a $d$-dimensional lattice space. Let $\bm{n}$ be an $d$-dimensional vector which characterize the spatial position of each
oscillator degree of freedom $(\hat{q}_{\bm{n}},\hat{p}_{\bm{ n}})$. The
extension of Eqs. (\ref{eq_partner_in_vacuum_dis_1})-(\ref{eq_partner_in_vacuum_dis_2}) to a $d$-dimensional lattice space can be
obtained by replacing $n$ into $\bm{n}$. Then, the continuum limit can be
taken. The partner formula for a $(d+1)$ dimensional quantum field is given
by 
\begin{eqnarray}
\hat{Q}_{A} &=&\left( \frac{\sqrt{1+g^{2}}}{2}\right) ^{1/2}\int d^{d}\bm{x}\left[ X_{A}(\bm{x})\hat{\phi}^{S}(\bm{x})+Y_{A}(\bm{x})\hat{\Pi}^{S}(\bm{x})\right]  \label{cont1} \\
\hat{P}_{A} &=&\left( \frac{\sqrt{1+g^{2}}}{2}\right) ^{1/2}\int d^{d}\bm{x}\left[ Z_{A}(\bm{x})\hat{\phi}^{S}(\bm{x})+W_{A}(\bm{x})\hat{\Pi}^{S}(\bm{x})\right] \\
\hat{Q}_{B} &=&\left( \frac{\sqrt{1+g^{2}}}{2}\right) ^{1/2}\int d^{d}\bm{x}\left[ X_{B}(\bm{x})\hat{\phi}^{S}(\bm{x})+Y_{B}(\bm{x})\hat{\Pi}^{S}(\bm{x})\right] \\
\hat{P}_{B} &=&\left( \frac{\sqrt{1+g^{2}}}{2}\right) ^{1/2}\int d^{d}\bm{x}\left[ Z_{B}(\bm{x})\hat{\phi}^{S}(\bm{x})+W_{B}(\bm{x})\hat{\Pi}^{S}(\bm{x})\right] ,
\end{eqnarray}
with the weighting functions of the partner $B$ written in terms of those of
the mode $A$ as follows: 
\begin{align}
X_{B}(\bm{x})& \equiv \frac{\sqrt{1+g^{2}}}{g}X_{A}(\bm{x})-\frac{2}{g}\int
d^{d}\bm{x'}\Delta _{p}(\bm{x}-\bm{x'})W_{A}(\bm{x'}),
\label{eq_partner_in_vacuum_1} \\
Y_{B}(\bm{x})& \equiv \frac{\sqrt{1+g^{2}}}{g}Y_{A}(\bm{x})+\frac{2}{g}\int
d^{d}\bm{x'}\Delta _{q}(\bm{x}-\bm{x'})Z_{A}(\bm{x'}), \\
Z_{B}(\bm{x})& \equiv -\frac{\sqrt{1+g^{2}}}{g}Z_{A}(\bm{x})-\frac{2}{g}\int
d^{d}\bm{x'}\Delta _{p}(\bm{x}-\bm{x'})Y_{A}(\bm{x'}), \\
W_{B}(\bm{x})& \equiv -\frac{\sqrt{1+g^{2}}}{g}W_{A}(\bm{x})+\frac{2}{g}\int
d^{d}\bm{x'}\Delta _{q}(\bm{x}-\bm{x'})X_{A}(\bm{x'}),
\label{eq_partner_in_vacuum_2}
\end{align}
where 
\begin{align}
\Delta _{q}\left( \bm{x}-\bm{x'}\right) & =\int \frac{d^{d}\bm{k}}{(2\pi
)^{d}}\frac{1}{2E_{\bm{k}}}e^{i\bm{k}\cdot (\bm{x}-\bm{x'})}, \\
\Delta _{p}\left( \bm{x}-\bm{x'}\right) & =\int \frac{d^{d}\bm{k}}{(2\pi
)^{d}}\frac{E_{\bm{k}}}{2}e^{i\bm{k}\cdot (\bm{x}-\bm{x'})}
\end{align}
with $E_{\bm{k}}\equiv \sqrt{\left|\bm{k}\right|^2+m^2}$.

\section{Partner mode in excited Gaussian states}\label{sec_partner_general}
Let us consider an $N$ harmonic oscillator system in a pure Gaussian state $\ket{\Psi}$. Here, we do not assume that $\ket{\Psi}$ is the ground state of the Hamiltonian of the system. It is known that there exists a second-order ``Hamiltonian'' $\hat{H}=\sum_{k=0}^{N-1} \omega _k \hat{b}_k^\dag \hat{b}_k$ whose ground state is $\ket{\Psi}$, where $\omega_k>0$, $b_k^\dag$ and $b_k$ are creation and annihilation operators, and $\Braket{\Psi|\hat{b}_k|\Psi}=0$ \cite{QCV}. Thus, if us fix a mode $A$ by
\begin{align}
 \hat{q}_A&=\sum_{n=1}^{N}\left(x(n) \hat{q}'_n+y(n)\hat{p}'_n\right),\quad
 \hat{p}_A= \sum_{n=1}^{N}\left(z(n) \hat{q}'_n+w(n)\hat{p}'_n\right),
\end{align}
where 
\begin{align}
 \hat{q}'_n=\sum_{k=0}^{N-1}\frac{1}{\sqrt{2\omega_k}}\left(\hat{b}_ku_k(n)+\hat{b}_k^\dag u_k(n)^*\right),\quad\hat{p}'_n=\frac{1}{i}\sum_{k=0}^{N-1}\sqrt{\frac{\omega_k}{2}}\left(\hat{b}_ku_{k}(n)-\hat{b}_k^\dag u_k(n)\right)\label{eq_qpprime},
\end{align}
the procedure to identify the partner mode $B$ presented in the previous section is applicable in a direct way. 

Now, let us derive a more general expression of the partner formula for an arbitrary Gaussian state $\ket{\Psi}$. Fix a complete set of canonical operators $\{(\hat{q}_n,\hat{p}_n)\}_{n=1}^N$ satisfying $[\hat{q}_n,\hat{p}_m]=i\delta_{nm}$, which is not necessarily assumed to be the same as that defined in Eqs. (\ref{eq_qp}) nor (\ref{eq_qpprime}). 
Without loss of generality, it is possible to assume $\Braket{\Psi|\hat{q}_n|\Psi}=0$ and $\Braket{\Psi|\hat{p}_n|\Psi}=0$ hold for all $n$ by shifting 
\begin{align}
 \hat{q}_n\to \hat{q}_n-\Braket{\Psi|\hat{q}_n|\Psi}, \quad \hat{p}_n\to \hat{p}_n-\Braket{\Psi|\hat{p}_n|\Psi}.
\end{align}
 Let us fix a mode $A$ characterized by weighting functions $\{(x(n),y(n),z(n),w(n))\}_{n=1}^N$ defined as
\begin{align}
 \hat{q}_A&=\sum_{n=1}^N\left(x(n)\hat{q}_n+y(n)\hat{p}_n\right)\equiv\bm{v}_A\sps{T}\hat{\bm{r}}, \\
\hat{p}_A&=\sum_{n=1}^N\left(z(n)\hat{q}_n+w(n)\hat{p}_n\right)\equiv\bm{u}_A\sps{T}\hat{\bm{r}},
\end{align}
where we have defined $\hat{\bm{r}}\equiv \left(\hat{q}_1,\hat{p}_1,\cdots,\hat{q}_{N},\hat{p}_{N}\right)^{\mathrm{T}}$ and $\bm{v}_A=(x(1),y(1),\cdots,x(N),y(N))^{\mathrm{T}},\bm{u}_A=(z(1),w(1),\cdots, z(N),w(N))^{\mathrm{T}}\in\mathbb{R}^{2N}$. Imposing $\left[\hat{q}_A,\hat{p}_A\right]=i$, the vectors must satisfy $\bm{v}_A\sps{T}\Omega \bm{u}_A=1$, where $\Omega$ is defined as
\begin{align}
 \Omega=\bigoplus_{n=1}^N
\begin{pmatrix}
 0&1\\
 -1&0 
\end{pmatrix}.
\end{align}
After an appropriate local symplectic transformation, it is possible to bring the set of operators to the standard form $\left(\hat{Q}_A,\hat{P}_A\right)\equiv\left(\bm{V}_A\sps{T}\hat{\bm{r}},\bm{U}_A\sps{T}\hat{\bm{r}}\right)$, such that
\begin{align}
 \begin{pmatrix}
  \Braket{\Psi|\hat{Q}_A^2|\Psi}&\rre\left(\Braket{\Psi|\hat{Q}_A\hat{P}_A|\Psi}\right)\\
  \rre\left(\Braket{\Psi|\hat{P}_A\hat{Q}_A|\Psi}\right)&\Braket{\Psi|\hat{P}_A^2|\Psi} 
 \end{pmatrix}
 =\frac{\sqrt{1+g^2}}{2}
\begin{pmatrix}
 1&0\\
 0&1 
\end{pmatrix}
\end{align}
holds, where $g\equiv \sqrt{4\left(\Braket{\Psi|\hat{q}_A^2|\Psi}\Braket{\Psi|\hat{p}_A^2|\Psi}-\rre\left(\Braket{\Psi|\hat{q_A}\hat{p}_A|\Psi}\right)\right)-1}$. This condition is equivalent to
\begin{align}
 \bm{V}_A\sps{T}M\bm{V}_A=\bm{U}_A\sps{T}M \bm{U}_A =\frac{\sqrt{1+g^2}}{2}, \quad \bm{V}_A\sps{T}M \bm{U}_A=\bm{U}_A\sps{T}M \bm{V}_A=0,  \label{eq_stdform_gen}
\end{align}
where we have defined the covariance matrix $M\equiv \rre\left(\Braket{\Psi|\hat{\bm{r}}\hat{\bm{r}}\sps{T}|\Psi}\right)$. From $\bm{v}_A\sps{T}\Omega \bm{u}_A=1$, we also have
\begin{align}
 \bm{V}_A\sps{T}\Omega \bm{U}_A=1.\label{eq_comconda_gen}
\end{align}
Now let us define another mode $B$ by $\left(\hat{Q}_B,\hat{P}_B\right)\equiv \left(\bm{V}_B\sps{T}\hat{r},\bm{U}_B\sps{T}\hat{r}\right)$, where
\begin{align}
 \bm{V}_B\sps{T}\Omega\bm{U}_B=1\label{eq_comcond_gen}
\end{align}
is assumed to be satisfied. From the locality condition and the purification condition, the mode $B$ is the partner of $A$ if and only if
\begin{align}
 \bm{V}_A\sps{T}\Omega\bm{V}_B &=\bm{V}_A\sps{T}\Omega\bm{ U}_B=\bm{U}_A\sps{T}\Omega \bm{V}_B=\bm{U}_A\sps{T}\Omega \bm{U}_B=0 
\end{align}
and
\begin{align}
 \bm{V}_A\sps{T}M\bm{V}_B&=-\bm{U}_A\sps{T}M\bm{U}_B=\frac{g}{2},\\
 \bm{V}_A\sps{T}M\bm{U}_B&=\bm{U}_A\sps{T}M\bm{V}_B=0 ,\\
 \bm{V}_B\sps{T}M\bm{V}_B&=\bm{U}_B\sps{T}M\bm{U}_B=\frac{\sqrt{1+g^2}}{2}, \\
 \bm{V}_B\sps{T}M\bm{U}_B&=0 \label{eq_purcond_gen}
\end{align}
hold. Since the partner mode is unique, if one could find $\bm{V}_B,\bm{U}_B\in\mathbb{R}^N$ satisfying equations (\ref{eq_comcond_gen})-(\ref{eq_purcond_gen}) under the constraints (\ref{eq_stdform_gen})-(\ref{eq_comconda_gen}), then the mode $B$ is the partner of $A$. From equations (\ref{eq_partner_in_vacuum_dis_1})-(\ref{eq_partner_in_vacuum_dis_2}), it is not hard to expect that
\begin{align}
 \bm{V}_B=\frac{\sqrt{1+g^2}}{g}\bm{V}_A-\frac{2}{g}\Omega M \bm{U}_A,\quad  \bm{U}_B=-\frac{\sqrt{1+g^2}}{g}\bm{U}_A-\frac{2}{g}\Omega M \bm{V}_A\label{eq_partner_excited}
\end{align}
satisfy the requirements. In fact, it can straightforwardly be verified by using $M\Omega M=\frac{1}{4}\Omega$. This identity always holds for pure Gaussian states $\ket{\Psi}$, which follows from the fact that there exists a symplectic matrix $S$ such that $M=\frac{1}{2}S S\sps{T}$ \cite{QCV}. 
In terms of weighting functions, Eq. (\ref{eq_partner_excited}) can be written as
\begin{align}
 X_B(n)&=\frac{\sqrt{1+g^2}}{g}X_A(n)-\frac{2}{g}\sum_{m=1}^N\left(\rre\left(\Braket{\Psi|\hat{p}_n\hat{q}_m|\Psi}\right)Z_A(m) +\Braket{\Psi|\hat{p}_n\hat{p}_m|\Psi}W_A(m)\right),\\
 Y_B(n)&=\frac{\sqrt{1+g^2}}{g}Y_A(n)+\frac{2}{g}\sum_{m=1}^N\left(\Braket{\Psi|\hat{q}_n\hat{q}_m|\Psi}Z_A(m)+\rre\left(\Braket{\Psi|\hat{q}_n\hat{p}_m|\Psi}\right)W_A(m)\right) ,\\
 Z_B(n)&=-\frac{\sqrt{1+g^2}}{g}Z_A(n)-\frac{2}{g}\sum_{m=1}^N\left(\rre\left(\Braket{\Psi|\hat{p}_n\hat{q}_m|\Psi}\right)X_A(m) +\Braket{\Psi|\hat{p}_n\hat{p}_m|\Psi}Y_A(m)\right), \\
 W_B(n)&=-\frac{\sqrt{1+g^2}}{g}W_A(n)+\frac{2}{g}\sum_{m=1}^N\left(\Braket{\Psi|\hat{q}_n\hat{q}_m|\Psi}X_A(m)+\rre\left(\Braket{\Psi|\hat{q}_n\hat{p}_m|\Psi}\right)Y_A(m)\right) .
\end{align}
These are the general partner formula, which can be used for any Gaussian state and any complete set of canonical operators.
As long as the continuum limit can be taken properly, we obtain the partner formula in the scalar field theory.
Especially, the partner formula for weighting functions of the field $\hat{\phi}(\bm{x})$ and its conjugate momentum $\hat{\Pi}(\bm{x})$ is given by
\begin{align}
 X_B(\bm{x})&=\frac{\sqrt{1+g^2}}{g}X_A(\bm{x})\nonumber\\
&\quad-\frac{2}{g}\int d^d\bm{y}\left(\rre\left(\Braket{\Psi|\hat{\Pi}(\bm{x})\hat{\phi}(\bm{y})|\Psi}\right)Z_A(\bm{y}) +\Braket{\Psi|\hat{\Pi}(\bm{x})\hat{\Pi}(\bm{y})|\Psi}W_A(\bm{y})\right),\label{eq_partner_wfs1}\\
 Y_B(\bm{x})&=\frac{\sqrt{1+g^2}}{g}Y_A(\bm{x})\nonumber\\
&\quad +\frac{2}{g}\int d^d\bm{y}\left(\Braket{\Psi|\hat{\phi}(\bm{x})\hat{\phi}(\bm{y})|\Psi}Z_A(\bm{y})+\rre\left(\Braket{\Psi|\hat{\phi}(\bm{x})\hat{\Pi}(\bm{y})|\Psi}\right)W_A(\bm{y})\right) ,\\
 Z_B(\bm{x})&=-\frac{\sqrt{1+g^2}}{g}Z_A(\bm{x})\nonumber\\
&\quad-\frac{2}{g}\int d^d\bm{y}\left(\rre\left(\Braket{\Psi|\hat{\Pi}(\bm{x})\hat{\phi}(\bm{y})|\Psi}\right)X_A(\bm{y}) +\Braket{\Psi|\hat{\Pi}(\bm{x})\hat{\Pi}(\bm{y})|\Psi}Y_A(\bm{y})\right), \\
 W_B(\bm{x})&=-\frac{\sqrt{1+g^2}}{g}W_A(\bm{x})\nonumber\\
&\quad +\frac{2}{g}\int d^d\bm{y}\left(\Braket{\Psi|\hat{\phi}(\bm{x})\hat{\phi}(\bm{y})|\Psi}X_A(\bm{y})+\rre\left(\Braket{\Psi|\hat{\phi}(\bm{x})\hat{\Pi}(\bm{y})|\Psi}\right)Y_A(\bm{y})\right) \label{eq_partner_wfs2}.
\end{align}
These are the partner formula written in terms of two-point functions.

\section{PARTNER MODE IN A CURVED SPACETIME}\label{sec_partner_CS}
By using the result obtained in the previous section, let us investigate
the memory effect in pairs of partners of free scalar field in a curved spacetime. The metric is denoted by $g_{\mu\nu}(x)$ whose signature is given by 
$(-,+,+,\cdots,+)$. Here $x$ denotes a point in the spacetime and Greek
indices run over $0,1,\cdots d$. Assuming the spacetime is globally
hyperbolic, it is possible to foliate the spacetime into a family of spatial
slices $\Sigma_\tau$, where $\tau$ denotes a continuous parameter which can
be regarded as time. For simplicity, we assume there are two regions, ``in'' and ``out''
region, where the spacetime becomes flat: 
\begin{equation}
ds^2=g_{\mu\nu}dx^{\mu}dx^{\nu}= 
\begin{cases}
& -dt^2+d\bm{x}^2\quad \text{(in the ``in'' region)}. \\ 
& -d\bar{t}^2+d\bm{\bar{x}}^2 \quad \text{(in the ``out'' region)}.
\end{cases}
\end{equation}
Here, $(t,\bm{x})$ and $(\bar{t},\bm{\bar{x}})$ are the coordinate system in
the ``in'' and ``out'' region, respectively. It should be stressed that we
have imposed no constraint on the metric in the intermediate region between
two flat regions as long as the spacetime is globally hyperbolic. An action for the free scalar field $\phi$ is given by 
\begin{align}
S=\int d^{d+1}x \sqrt{-g(x)}\frac{1}{2}\left(-g^{\mu\nu}\partial_{\mu}\phi\partial_{\nu}\phi-\left(m(x)^2+\xi R(x)\right)\phi^2\right),
\end{align}
where $m(x)$ is the mass of the scalar field which may depend on the
position $x$, $R(x)$ is the Ricci scalar of the spacetime and $\xi$
characterize the coupling between the scalar field and the gravitational
field. Adopting the Heisenberg picture, the equation of motion is given by the Klein-Gordon equation $\left(\Box+m(x)^2+\xi R(x)\right)\hat{\phi}(x)=0$, where $\Box \phi\equiv 
\frac{1}{\sqrt{-g}}\partial_{\mu}\left(\sqrt{-g}g^{\mu\nu}\partial_\nu \hat{\phi}\right)$. The conjugate momentum is given as 
\begin{equation}
\hat{\Pi}(x)=-\sqrt{-g}g^{\tau\mu}\partial_\mu \hat{\phi}(x)= 
\begin{cases}
& \partial_{t} \hat{\phi}(t,\bm{x})\quad \text{(in the ``in'' region)}. \\ 
& \partial_{\bar{t}}\hat{\phi}(\bar{t},\bar{\bm{x}})\quad \text{(in the
``out'' region)}.
\end{cases}
\end{equation}
In the flat region, the Ricci scalar vanishes. Let us assume $m(x)$ becomes
constant in the flat regions as follows: 
\begin{equation}
m(x)= 
\begin{cases}
& m\quad \text{(in the ``in'' region)}. \\ 
& \bar{m}\quad \text{(in the ``out'' region)}.
\end{cases}
\end{equation}
Then, there are two sets of solutions for the equation of motion which
satisfy the following conditions: 
\begin{align}
u_{\bm{k}}(t,\bm{x})&= \frac{1}{\sqrt{(2\pi)^d 2E_{\bm{k}}}}e^{i(\bm{k}\cdot\bm{x}-E_{\bm{k}}t)}\quad \text{(in the ``in'' region)}, \\
\bar{u}_{\bm{k}}(\bar{t},\bm{\bar{x}})&= \frac{1}{\sqrt{(2\pi)^d 2\bar{E}_{\bm{k}}}}e^{i(\bm{k}\cdot \bm{\bar{x}}-\bar{E}_{\bm{k}}\bar{t})}\quad \text{(in the ``out'' region)},
\end{align}
where $E_{\bm{k}}\equiv \sqrt{\bm{k}^2+m^2}$ and $\bar{E}_{\bm{k}}\equiv 
\sqrt{\bm{k}^2+\bar{m}^2}$ are energies for the field with momentum $\bm{k}$
in ``in'' region and ``out'' region, respectively. The normalization
constants are chosen to satisfy $(u_{\bm{k}},u_{\bm{k}^{\prime }})=(\bar{u}_{\bm{k}},\bar{u}_{\bm{k}^{\prime }})=\delta^{(d)}(\bm{k}-\bm{k}^{\prime })$,
where we have introduced the inner product of functions $f_1,f_2$ as 
\begin{align}
(f_1,f_2)\equiv -i \int_{\Sigma_\tau} d\Sigma^\mu\sqrt{g_{\Sigma}(x)}f_1(x)\overleftrightarrow{\partial_\mu}f_2(x)^*\equiv -i \int_{\Sigma_\tau}
d\Sigma^\mu\sqrt{g_{\Sigma}(x)}\left(f_1(x)\partial_\mu f_2(x)^*-
f_2(x)^*\partial_\mu f_1(x)\right).
\end{align}
Here, $g_{\Sigma}$ is the determinant of the induced metric on the time
slice $\Sigma_\tau$, $d\Sigma^\mu\equiv n^\mu d\Sigma$ with a unit normal
vector $n^\mu$ and the volume element $d\Sigma$ of the spatial slice $\Sigma_\tau$. It
should be noted that for solutions $f_1,f_2$ of the equation of motion, it
can be shown that the inner product is independent of the choice of $\Sigma_\tau$. For each complete set of solutions, a set of creation and
annihilation operators is introduced in the following way: 
\begin{align}
\hat{\phi}^{\mathrm{H}} (x)=\int d^d \bm{k}\left(\hat{a}_{\bm{k}}u_{\bm{k}}(x)+\hat{a}_{\bm{k}}^\dag u_{\bm{k}}(x)^*\right)=\int d^d \bm{k}\left(\hat{\bar{a}}_{\bm{k}}\bar{u}_{\bm{k}}(x)+\hat{\bar{a}}_{\bm{k}}^\dag \bar{u}_{\bm{k}}(x)^*\right),
\end{align}
where the superscript H of $\hat{\phi}$ is added to emphasize we adopt the
Heisenberg picture. They are related with each other through 
\begin{align}
\hat{a}_{\bm{k}}=\left(\phi,u_{\bm{k}}\right)=\int d^d \bm{k}^{\prime
}\left(\alpha_{\bm{k}^{\prime }\bm{k}}\hat{\bar{a}}_{\bm{k}^{\prime
}}+\beta_{\bm{k}^{\prime }\bm{k}}\hat{\bar{a}}_{\bm{k}^{\prime
}}^\dag\right),
\end{align}
where the Bogoliubov coefficients are defined by 
\begin{align}
\alpha_{\bm{k}^{\prime }\bm{k}}\equiv\left(\bar{u}_{\bm{k}^{\prime }},u_{\bm{k}}\right),\quad \beta_{\bm{k}^{\prime }\bm{k}}\equiv\left(\bar{u}_{\bm{k}^{\prime }}^*,u_{\bm{k}}\right).  \label{Bogo_def}
\end{align}
The inverse transformation is given by 
\begin{align}
\hat{\bar{a}}_{\bm{k}}=\int d^d \bm{k}^{\prime }\left(\alpha_{\bm{k}\bm{k}^{\prime }}^*\hat{a}_{\bm{k}^{\prime }}-\beta_{\bm{k}\bm{k}^{\prime }}^*\hat{a}_{\bm{k}^{\prime }}^\dag\right).
\end{align}

Since the formula obtained in the previous section is applicable for any Gaussian state and any complete set of canonical operators, it is possible to obtain the partner even when we are working in the Heisenberg picture. As an example, at $\tau=\bar{t}$ in the ``out'' region, let us consider a mode $A$ characterized by
\begin{align}
 \hat{q}_A\sps{H}&=\int d^d\bar{\bm{x}}\left(x_A(\bar{\bm{x}})\hat{\phi}\sps{H}(\bar{t},\bar{\bm{x}})+y_A(\bar{\bm{x}})\hat{\Pi}\sps{H}(\bar{t},\bar{\bm{x}})\right),\\
 \hat{p}_A\sps{H}&=\int d^d\bar{\bm{x}}\left(z_A(\bar{\bm{x}})\hat{\phi}\sps{H}(\bar{t},\bar{\bm{x}})+w_A(\bar{\bm{x}})\hat{\Pi}\sps{H}(\bar{t},\bar{\bm{x}})\right) ,
\end{align}
satisfying $[\hat{q}_A\sps{H},\hat{p}_A\sps{H}]=i$.
After an appropriate local symplectic transformation, the canonical operators reduce to the standard form 
\begin{align}
 \hat{Q}_A\sps{H}&=\int d^d\bar{\bm{x}}\left(X_A(\bar{\bm{x}})\hat{\phi}\sps{H}(\bar{t},\bar{\bm{x}})+Y_A(\bar{\bm{x}})\Pi\sps{H}(\bar{t},\bar{\bm{x}})\right),\\
 \hat{P}_A\sps{H}&=\int d^d\bar{\bm{x}}\left(Z_A(\bar{\bm{x}})\hat{\phi}\sps{H}(\bar{t},\bar{\bm{x}})+W_A(\bar{\bm{x}})\Pi\sps{H}(\bar{t},\bar{\bm{x}})\right) ,
\end{align}
which satisfy
\begin{align}
 \begin{pmatrix}
  \Braket{\Psi|\left(\hat{Q}_A\sps{H}\right)^2|\Psi} & \rre\left(\Braket{\Psi|\hat{Q}\sps{H}_A\hat{P}\sps{H}_A|\Psi}\right)\\
  \rre\left(\Braket{\Psi|\hat{P}\sps{H}_A\hat{Q}\sps{H}_A|\Psi}\right)&\Braket{\Psi|\left(\hat{P}\sps{H}_A\right)^2|\Psi} 
 \end{pmatrix}
=\frac{\sqrt{1+g^2}}{2}
\begin{pmatrix}
 1&0\\
 0&1 
\end{pmatrix},
\end{align}
where $g\equiv\sqrt{4\left(\Braket{\Psi|\left(\hat{q}_A\sps{H}\right)^2|\Psi}\Braket{\Psi|\left(\hat{p}_A\sps{H}\right)^2|\Psi}-\left(\rre\left(\Braket{\Psi|q_A\sps{H}p_A\sps{H}|\Psi}\right)\right)\right)-1}$ and $\ket{\Psi}$ is a Gaussian state of the field, which is typically taken as the vacuum state in the ``in'' region.
The weighting functions for the partner $B$ are given by
\begin{align}
 X_B(\bm{\bar{x}})&=\frac{\sqrt{1+g^2}}{g}X_A(\bm{\bar{x}})\nonumber\\
&\quad-\frac{2}{g}\int d^d\bm{\bar{y}}\left(\rre\left(\Braket{\Psi|\hat{\Pi}\sps{H}(\bar{t},\bm{\bar{x}})\hat{\phi}\sps{H}(\bar{t},\bm{\bar{y}})|\Psi}\right)Z_A(\bm{\bar{y}}) +\Braket{\Psi|\hat{\Pi}\sps{H}(\bar{t},\bm{\bar{x}})\hat{\Pi}\sps{H}(\bar{t},\bm{\bar{y}})|\Psi}W_A(\bm{\bar{y}})\right),\label{eq_partner_hei1}\\
 Y_B(\bm{\bar{x}})&=\frac{\sqrt{1+g^2}}{g}Y_A(\bm{\bar{x}})\nonumber\\
&\quad +\frac{2}{g}\int d^d\bm{\bar{y}}\left(\Braket{\Psi|\hat{\phi}\sps{H}(\bar{t},\bm{\bar{x}})\hat{\phi}\sps{H}(\bar{t},\bm{\bar{y}})|\Psi}Z_A(\bm{\bar{y}})+\rre\left(\Braket{\Psi|\hat{\phi}\sps{H}(\bar{t},\bm{\bar{x}})\hat{\Pi}\sps{H}(\bar{t},\bm{\bar{y}})|\Psi}\right)W_A(\bm{\bar{y}})\right) ,\\
 Z_B(\bm{\bar{x}})&=-\frac{\sqrt{1+g^2}}{g}Z_A(\bm{\bar{x}})\nonumber\\
&\quad-\frac{2}{g}\int d^d\bm{\bar{y}}\left(\rre\left(\Braket{\Psi|\hat{\Pi}\sps{H}(\bar{t},\bm{\bar{x}})\hat{\phi}\sps{H}(\bar{t},\bm{\bar{y}})|\Psi}\right)X_A(\bm{\bar{y}}) +\Braket{\Psi|\hat{\Pi}\sps{H}(\bar{t},\bm{\bar{x}})\hat{\Pi}\sps{H}(\bar{t},\bm{\bar{y}})|\Psi}Y_A(\bm{\bar{y}})\right), \\
 W_B(\bm{\bar{x}})&=-\frac{\sqrt{1+g^2}}{g}W_A(\bm{\bar{x}})\nonumber\\
&\quad +\frac{2}{g}\int d^d\bm{\bar{y}}\left(\Braket{\Psi|\hat{\phi}\sps{H}(\bar{t},\bm{\bar{x}})\hat{\phi}\sps{H}(\bar{t},\bm{\bar{y}})|\Psi}X_A(\bm{\bar{y}})+\rre\left(\Braket{\Psi|\hat{\phi}\sps{H}(\bar{t},\bm{\bar{x}})\hat{\Pi}\sps{H}(\bar{t},\bm{\bar{y}})|\Psi}\right)Y_A(\bm{\bar{y}})\right).\label{eq_partner_hei2}
\end{align}

Now, consider a situation in which an experimenter prepares an Unruh-DeWitt
particle detector at $\bar{t}=\bar{t}_{\mathrm{obs.}} $ in the ``out''
region, which couples with a mode of the field to read out quantum
information imprinted in the field. To perform such a protocol, one has to
consider an interaction between the field and an external device. Therefore,
it is useful to obtain a partner formula based on the Schr\"odinger picture. 
We want the partner of a mode $A$ whose canonical variables are defined by 
\begin{align}
\hat{q}_A^{\mathrm{S}} &=\int d^d \bm{\bar{x}}\left(x_A(\bm{\bar{x}})\hat{\phi}^{\mathrm{S}} (\bm{\bar{x}})+y_A(\bm{\bar{x}})\hat{\Pi}^{\mathrm{S}} (\bm{\bar{x}})\right), \\
\hat{p}_A^{\mathrm{S}} & =\int d^d \bm{\bar{x}}\left(z_A(\bm{\bar{x}})\hat{\phi}^{\mathrm{S}} (\bm{\bar{x}})+w_A(\bm{\bar{x}})\hat{\Pi}^{\mathrm{S}} (\bm{\bar{x}})\right),
\end{align}
where the superscript S of $\hat{\phi}$ and $\hat{\Pi}$ are added to
emphasize that we adopt the Sch\"odinger picture. Since the pair of canonical
variables $(\hat{q}_A^{\mathrm{S}} ,\hat{p}_A^{\mathrm{S}} )$ must satisfy 
\begin{align}
\left[\hat{q}_A^{\mathrm{S}} ,\hat{p}_A^{\mathrm{S}} \right]=i,
\end{align}
we have the following constraint: 
\begin{align}
\int d^d \bm{\bar{x}} \left(x_A(\bm{\bar{x}})w_{A}(\bm{\bar{x}})-y_A(\bm{\bar{x}})z_A(\bm{\bar{x}})\right)=1,
\end{align}
where we have used the canonical commutation relationship of the field and
its conjugate momentum. Assuming the system is in the vacuum state $\ket{0}$
at $\tau=t_0$ in the ``in'' region, it evolves into $\ket{\psi(\bar{t})}=U(\bar{t},t_0)\ket{0}$ in the ``out'' region, where 
\begin{align}
U(\bar{t},t_0)\equiv \mathcal{T}\exp{\left(-i\int_{t_0}^{\bar{t}}d\tau\int_{\Sigma_\tau } d^d \bm{x}\mathcal{H}
\right)}\label{eq_unitary_evol}
\end{align}
is the unitary evolution operator. Here, the Hamiltonian density $\mathcal{H}$ is defined by
\begin{align}
 \mathcal{H}\equiv :\hat{\Pi}\partial_\tau\hat{\phi} -\mathcal{L}(\hat{\phi}):.
\end{align}
The excited state $\ket{\psi(\bar{t})}$ is a Gaussian state since the initial state $\ket{0}$ is a Gaussian state
and the Hamiltonian is bi-linear. 
Under the assumption that Eq. (\ref{eq_unitary_evol}) is well defined, the Heisenberg operators and the Schr\"odinger operators are related through $\hat{\phi}\sps{H}(\bar{t},\bm{\bar{x}})=U(\bar{t},t_0)^\dag\hat{\phi}\sps{S}(\bm{x})U(\bar{t},t_0)$ and $\hat{\Pi}\sps{H}(\bar{t},\bm{\bar{x}})=U(\bar{t},t_0)^\dag\hat{\Pi}\sps{S}(\bm{x})U(\bar{t},t_0)$. Therefore, the partner mode $B$ is characterized by
\begin{align}
\hat{Q}_B^{\mathrm{S}} &=\int d^d \bm{\bar{x}}\left(X_B(\bm{\bar{x}})\hat{\phi}^{\mathrm{S}} (\bm{\bar{x}})+Y_B(\bm{\bar{x}})\hat{\Pi}^{\mathrm{S}} (\bm{\bar{x}})\right), \\
\hat{P}_B^{\mathrm{S}} & =\int d^d \bm{\bar{x}}\left(Z_B(\bm{\bar{x}})\hat{\phi}^{\mathrm{S}} (\bm{\bar{x}})+W_B(\bm{\bar{x}})\hat{\Pi}^{\mathrm{S}} (\bm{\bar{x}})\right),
\end{align}
where the weighting functions are defined in Eqs. (\ref{eq_partner_hei1})-(\ref{eq_partner_hei2}) with $\ket{\Psi}=\ket{0}$. In terms of the Bogoliubov coefficients, the second moments of the field and its conjugate momentum are calculated as
\begin{align}
 &\Braket{0|\hat{\phi}\sps{H}(\bar{t},\bm{\bar{x}})\hat{\phi}\sps{H}(\bar{t},\bar{\bm{y}})|0}\nonumber\\
&=\int d^d\bm{k}d^d\bm{k}'d^d\bm{p}\left(\bar{u}_{\bm{k}}(\bar{t},\bm{\bar{x}})\alpha_{\bm{k}\bm{p}}^*-\bar{u}^*_{\bm{k}}(\bar{t},\bm{\bar{x}})\beta_{\bm{k}\bm{p}}\right)\left(\bar{u}_{\bm{k}'}^*(\bar{t},\bm{\bar{y}})\alpha_{\bm{k}'\bm{p}}-\bar{u}_{\bm{k}'}(\bar{t},\bm{\bar{y}})\beta_{\bm{k}'\bm{p}}^*\right),\\
& \rre\left(\Braket{0|\hat{\phi}\sps{H}(\bar{t},\bm{\bar{x}})\hat{\Pi}\sps{H}(\bar{t},\bm{\bar{y}})|0}\right)\nonumber\\
& =\int d^d\bm{k}d^d\bm{k}'d^d\bm{p}\rre\left[\left(\bar{u}_{\bm{k}}(\bar{t},\bm{\bar{x}})\alpha_{\bm{k}\bm{p}}^*-\bar{u}^*_{\bm{k}}(\bar{t},\bm{\bar{x}})\beta_{\bm{k}\bm{p}}\right)(i\bar{E}_{\bm{k}'})\left(\bar{u}_{\bm{k}'}^*(\bar{t},\bm{\bar{y}})\alpha_{\bm{k}'\bm{p}}+\bar{u}_{\bm{k}'}(\bar{t},\bm{\bar{y}})\beta_{\bm{k}'\bm{p}}^*\right)\right]\nonumber\\
 & =-\int d^d\bm{k}d^d\bm{k}'d^d\bm{p}\rim\left[\left(\bar{u}_{\bm{k}}(\bar{t},\bm{\bar{x}})\alpha_{\bm{k}\bm{p}}^*-\bar{u}^*_{\bm{k}}(\bar{t},\bm{\bar{x}})\beta_{\bm{k}\bm{p}}\right)\bar{E}_{\bm{k}'}\left(\bar{u}_{\bm{k}'}^*(\bar{t},\bm{\bar{y}})\alpha_{\bm{k}'\bm{p}}+\bar{u}_{\bm{k}'}(\bar{t},\bm{\bar{y}})\beta_{\bm{k}'\bm{p}}^*\right)\right],\\
 &\Braket{0|\hat{\Pi}\sps{H}(\bar{t},\bm{\bar{x}})\hat{\Pi}\sps{H}(\bar{t},\bm{\bar{y}})|0} \nonumber\\
 &=\int d^d\bm{k}d^d\bm{k}'d^d\bm{p}\bar{E}_{\bm{k}}\bar{E}_{\bm{k}'}\left(\bar{u}_{\bm{k}}(\bar{t},\bm{\bar{x}})\alpha_{\bm{k}\bm{p}}^*+\bar{u}^*_{\bm{k}}(\bar{t},\bm{\bar{x}})\beta_{\bm{k}\bm{p}}\right)\left(\bar{u}_{\bm{k}'}^*(\bar{t},\bm{\bar{y}})\alpha_{\bm{k}'\bm{p}}+\bar{u}_{\bm{k}'}(\bar{t},\bm{\bar{y}})\beta_{\bm{k}'\bm{p}}^*\right).
\end{align}
It should be noted that the second moments of the mode $A$, which are needed to obtain $g$, are calculated from the above moments. For example,
\begin{align}
& \Braket{0|\left(\hat{q}_A\sps{H}\right)^2|0}\nonumber\\
 &= \int d^d\bm{\bar{x}}d^d\bm{\bar{y}}\left(x_A(\bm{\bar{x}})\Braket{0|\hat{\phi}\sps{H}(\bar{t},\bm{\bar{x}})\hat{\phi}\sps{H}(\bar{t},\bar{\bm{y}})|0}x_A(\bm{\bar{y}})\right.\nonumber\\
&\left.\quad +2x_A(\bm{\bar{x}})\rre\left(\Braket{0|\hat{\phi}\sps{H}(\bar{t},\bm{\bar{x}})\hat{\Pi}\sps{H}(\bar{t},\bm{\bar{y}})|0}\right)y_A(\bm{\bar{y}})+y_A(\bm{\bar{x}})\Braket{0|\hat{\Pi}\sps{H}(\bar{t},\bm{\bar{x}})\hat{\Pi}\sps{H}(\bar{t},\bm{\bar{y}})|0}y_A(\bm{\bar{y}})\right).
\end{align}

Even after the spacetime becomes static in the \textquotedblleft
out\textquotedblright\ region, the partner has non-trivial dynamics in
general. That is, if the experiment is performed at $\bar{t}=\bar{t}_{\mathrm{obs.}}^{\prime }(\neq \bar{t}_{\mathrm{obs.}})$, the weighting
functions of the partner $B$ will be different from those at $\bar{t}=\bar{t}_{\mathrm{obs.}}$. This is natural because the partner is capable of
evolving in time due to the free evolution of the field. Our formula enables
us to identify the unique partner mode $B$, once $\bar{t}_{\mathrm{obs.}}$
is specified in the ``out'' region.

The partner formula on
Eqs. (\ref{eq_partner_hei1})-(\ref{eq_partner_hei2}) are simplified when the Bogoliubov coefficients
satisfy the following conditions: 
\begin{equation*}
\alpha _{\bm{k}\bm{k}'}=|r|^{-d/2}\alpha _{\bm{k}}\delta ^{(d)}\left( \bm{k'}-r^{-1}\bm{k}\right) ,\quad \beta _{\bm{k}\bm{k}'}=|r|^{-d/2}\beta _{\bm{k}}\delta ^{(d)}\left( \bm{k'}+r^{-1}\bm{k}\right) ,\quad \alpha _{-\bm{k}}=\alpha
_{\bm{k}},\quad \beta _{-\bm{k}}=\beta _{\bm{k}}
\end{equation*}
for a nonzero real number $r$ with spatial homogeneity. 
The normalization condition 
\begin{align}
\int d^d\bm{k}^{\prime \prime }\left(\alpha_{\bm{k}\bm{k}^{\prime \prime
}}\alpha_{\bm{k}^{\prime }\bm{k}^{\prime \prime }}^*-\beta_{\bm{k}\bm{k}^{\prime \prime }}\beta_{\bm{k}^{\prime }\bm{k}^{\prime \prime
}}^*\right)=\delta^{(d)}\left(\bm{k}-\bm{k}^{\prime }\right)
\end{align}
is equivalent to 
\begin{align}
\left|\alpha_{\bm{k}}\right|^2-\left|\beta_{\bm{k}}\right|^2=1.
\end{align}
For this case, if an
experiment is performed in the late time, i.e., $\bar{t}_{\mathrm{obs.}}\rightarrow \infty $, the entanglement entropy and the weighting functions
of the partner $B$ become independent of $\bar{t}_{\mathrm{obs.}}$. In addition, it can be shown that partner $B$
only stores information related to particle creation effects $\left\vert
\beta _{\bm{k}}\right\vert ^{2}$. This fact implies that the entanglement partners contain long-lasting memory of the dynamics of evolution. For proof, let us show that the state itself becomes independent of $\bar{t}\sub{obs.}$ and depend only on $|\beta_{\bm{k}}|^2$ in the limit of $\bar{t}\sub{obs.}\to\infty$. This properties can be checked form the following calculations on the elements of the covariant matrix in the limit of $\bar{t}\to\infty$:
\begin{align}
& \Braket{0|\hat{\phi}\sps{H}(\bar{t},\bm{\bar{x}})\hat{\phi}\sps{H}(\bar{t},\bm{\bar{y}})|0}\nonumber\\
&=\int d^d\bm{k}d^d\bm{p}d^d\bm{k}'d^d\bm{p}'\left(\bar{u}_{\bm{k}}(\bar{t},\bm{\bar{x}})\alpha_{\bm{k}\bm{p}}^*-\bar{u}_{\bm{k}}^*(\bar{t},\bm{\bm{x}})\beta_{\bm{k}\bm{p}}\right)\left(\bar{u}_{\bm{k}'}^*(\bar{t},\bm{\bar{y}})\alpha_{\bm{k}'\bm{p}'}-\bar{u}_{\bm{k}'}(\bar{t},\bm{\bar{y}})\beta_{\bm{k}'\bm{p}'}^*\right)\delta^{(d)}(\bm{p}-\bm{p}')\nonumber\\
 &=\int d^d\bm{k} \left(\bar{u}_{\bm{k}}(\bar{t},\bm{\bar{x}})\alpha_{\bm{k}}^*-\bar{u}_{-\bm{k}}^*(\bar{t},\bm{\bar{x}})\beta_{-\bm{k}}\right)\left(\bar{u}_{\bm{k}}^*(\bar{t},\bm{\bar{y}})\alpha_{\bm{k}}-\bar{u}_{-\bm{k}}(\bar{t},\bm{\bar{y}})\beta_{-\bm{k}}^*\right)\nonumber\\
 &\to \int \frac{d^d\bm{k}}{(2\pi)^d}\frac{1}{2\bar{E}_{\bm{k}}}\left(1+2\left|\beta_{\bm{k}}\right|^2\right)e^{i\bm{k}\cdot(\bm{\bar{x}}-\bm{\bar{y}})},\\
 &\rre\left(\Braket{0|\hat{\phi}\sps{H}(\bar{t},\bm{\bar{x}})\hat{\Pi}\sps{H}(\bar{t},\bm{\bar{x}}')|0} \right)\to 0,\\
 &\Braket{0|\hat{\Pi}\sps{H}(\bar{t},\bm{\bar{x}})\hat{\Pi}\sps{H}(\bar{t},\bm{\bar{x}}')|0} \to\int \frac{d^d\bm{k}}{(2\pi)^2}\frac{\bar{E}_{\bm{k}}}{2}(1+2|\beta_{\bm{k}}|^2)e^{i\bm{k}\cdot(\bm{x}-\bm{x}')}, 
\end{align}
where we have used the Riemann--Lebesgue lemma, which claims that the
Fourier coefficient will vanish for high frequency modes. More precisely,
for an $L^1$ function $f$, it holds that 
\begin{align}
\tilde{f}(z)\equiv \int_{\mathbb{R}^d}dx f(x)e^{izx} \to 0\quad
(|z|\to\infty).
\end{align}
A rough proof for the one-dimensional case is given by the integration by
parts as follows: 
\begin{align}
\left|\tilde{f}(z)\right|=\left|-\int dx\frac{1}{iz}\frac{d}{dx}f(x)e^{izx}
\right|\leq \frac{1}{\left|z\right|}\int dx \left|\frac{d}{dx}f(x)\right|
\to 0\quad (|z|\to\infty).
\end{align}

As an example of our partner formula, let us consider a $(1+1)$-dimensional expanding universe model \cite{BD1,BD2} whose metric is defined
by 
\begin{equation*}
ds^{2}=\left( a+b\tanh \left( \rho \eta \right) \right) \left( -d\eta
^{2}+d\xi ^{2}\right) ,
\end{equation*}
where $\eta $ is the conformal time. Since the mass is independent of the
position in the spacetime in this model, $m=\bar{m}$. There are two
asymptotic regions $\eta \rightarrow -\infty $ and $\eta \rightarrow \infty $
where the spacetime becomes flat. We assume that the field is in the vacuum state in the ``in'' region. Assuming the periodic boundary condition: $\hat{\phi}(\eta,\xi+L)=\hat{\phi}(\eta,\xi)$, the unitary evolution matrix in Eq. (\ref{eq_unitary_evol}) exists. In the limit of $L\to\infty$, the dispersion relations are given by $\bar{E}_{k}=E_{k}=\sqrt{k^{2}+m^{2}}$. By using the result in \cite{BD1,BD2}, the Bogoliubov coefficients are obtained as 
\begin{equation*}
\alpha _{kk^{\prime }}=r^{-1/2}\alpha _{k}\delta \left( k^{\prime}-r^{-1}k\right)
,\quad \beta _{kk^{\prime }}=r^{-1/2}\beta _{k}\delta \left( k^{\prime}-r^{-1}k\right) ,
\end{equation*}
where $r\equiv \sqrt{\frac{a-b}{a+b}}$, $\alpha _{k}\equiv \tilde{\alpha}_{\sqrt{a+b}k}^{\ast }$ and $\beta _{k}\equiv -\tilde{\beta}_{-\sqrt{a+b}k}$
with 
\begin{align}
\tilde{\alpha}_{k}& \equiv \sqrt{\frac{\bar{\omega}_{k}}{\omega _{k}}}\frac{\Gamma \left( 1-i\omega _{k}/\rho \right) \Gamma \left( -i\bar{\omega}_{k}/\rho \right) }{\Gamma \left( -i\omega _{+}/\rho \right) \Gamma \left(
1-i\omega _{+}/\rho \right) } \label{eq_bogo_example1}\\
\tilde{\beta}_{k}& \equiv \sqrt{\frac{\bar{\omega}_{k}}{\omega _{k}}}\frac{\Gamma \left( 1-i\omega _{k}/\rho \right) \Gamma \left( i\bar{\omega}_{k}/\rho \right) }{\Gamma \left( i\omega _{-}/\rho \right) \Gamma \left(1+i\omega _{-}/\rho \right) }.\label{eq_bogo_example2}
\end{align}
Here, we have defined $\omega _{k}\equiv \sqrt{k^{2}+(a-b)m^{2}}$, $\bar{\omega}_{k}\equiv \sqrt{k^{2}+(a+b)m^{2}}$ and $\omega _{\pm }\equiv \frac{1}{2}\left( \bar{\omega}_{k}\pm \omega _{k}\right) $. Eqs. (\ref{eq_bogo_example1}) and (\ref{eq_bogo_example2}) are good approximation for finite and large $L$, where the evolution is unitary. In this example, $r$ is
related to the ratio between final and initial conformal factors with
respect to the conformal time $\eta $. In the limit of $\bar{t}_{\mathrm{obs.}}\rightarrow \infty $, the only contribution to the partner's weighting
functions comes from the particle creation rate $\left\vert \beta _{\bm{k}}\right\vert ^{2}$.

The parameter $g$ fixes the entanglement entropy as $S_{\mathrm{EE}} =\sqrt{1+g^2}\ln{\left(\frac{1}{g}\left(\sqrt{1+g^2}+1\right)\right)}+\ln{\left(\frac{g}{2}\right)}$, where $g$ is determined by
\begin{align}
 g^2=4\left(\Braket{0|\left(\hat{q}_A\sps{H}\right)^2|0}\Braket{0|\left(\hat{p}_A\sps{H}\right)^2|0}-\rre\left(\Braket{0|\hat{q}_A\sps{H}\hat{p}_A\sps{H}|0}\right)\right)-1.
\end{align}
The two-point functions of partner mode $B$ satisfy
\begin{align}
 g^2=4\Braket{0|\left(\hat{Q}_B\sps{H}\right)^2|0}\Braket{0|\left(\hat{P}_B\sps{H}\right)^2|0}-1.
\end{align}
Thus, the elements of covariance matrix such as
\begin{align}
& \Braket{0|\left(\hat{Q}_B\sps{H}\right)^2|0}\nonumber\\
 &=\int d^d\bm{\bar{x}}d^d\bm{\bar{y}}\left(X_B(\bar{\bm{x}})\Braket{0|\hat{\phi}\sps{H}(\bar{t},\bm{x})\hat{\phi}\sps{H}(\bar{t},\bar{\bm{y}})|0}X_B(\bar{\bm{y}})\right.\nonumber\\
&\left.\quad +2X_B(\bar{\bm{x}})\rre\left(\Braket{0|\hat{\phi}\sps{H}(\bar{t},\bar{\bm{x}})\hat{\Pi}\sps{H}(\bar{t},\bar{\bm{y}})|0}\right)Y_B(\bar{\bm{y}})+Y_B(\bar{\bm{x}})\Braket{0|\hat{\Pi}\sps{H}(\bar{t},\bar{\bm{x}})\hat{\Pi}\sps{H}(\bar{t},\bar{\bm{y}})|0}Y_B(\bar{\bm{y}})\right) 
\end{align}
are integrable.

\begin{figure}[]
\centering
\includegraphics[scale=1]{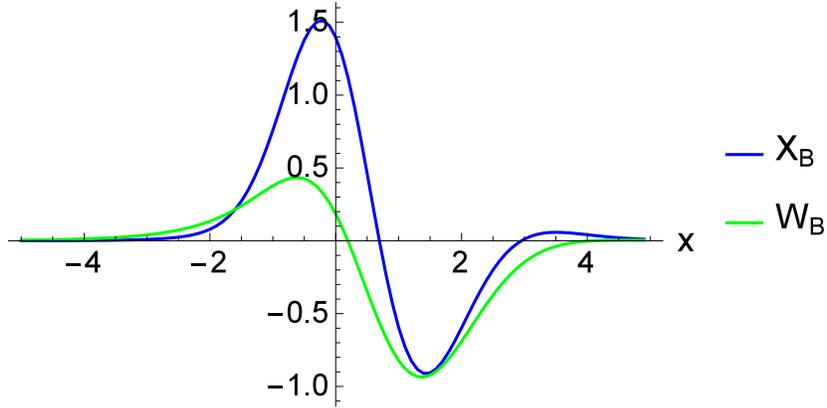}
\caption{Partner mode $B$ associated to the Gaussian mode $A$ in figure 
\protect\ref{GaussianA} when there is no expansion of the universe ($\protect\rho=0$). The mass of the scalar field was taken to be $m=1$.}
\label{GaussianBrho0}
\end{figure}

\begin{figure}[]
\centering
\includegraphics[scale=1]{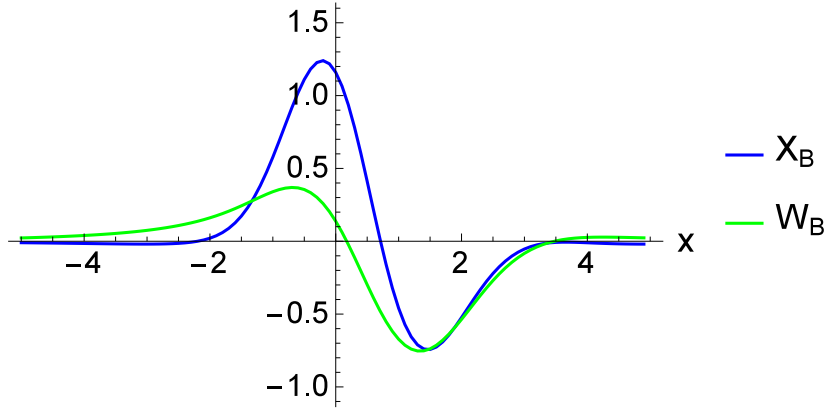}
\caption{Partner mode $B$ associated to the Gaussian mode $A$ in figure 
\protect\ref{GaussianA} when the expansion rate $\protect\rho=10$. In this
model, the universe starts from a size of ($a-b=0.5$) in the remote past and
ends with a size ($a+b=2.5$) in the remote future. The mass of the scalar
field was taken to be $m=1$.}
\label{GaussianBrho10}
\end{figure}

\begin{figure}[]
\centering
\includegraphics[scale=0.70]{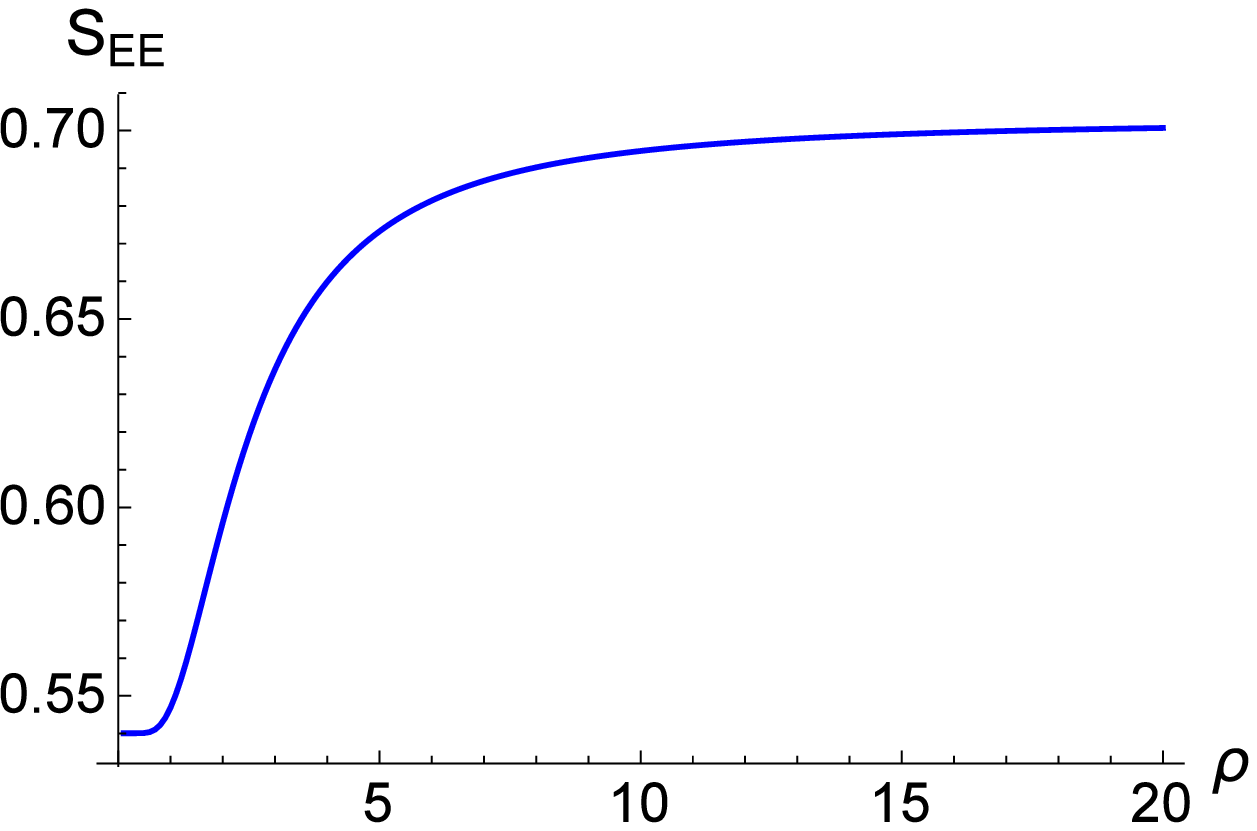}
\caption{Entanglement Entropy $S_{\mathrm{EE}} $ associated to the Gaussian
mode $A$ in figure \protect\ref{GaussianA} as a function of the universe
expansion rate $\protect\rho$. The same values as in figure \protect\ref{GaussianBrho10} are adopted for $a,b$ and $m$.}
\label{GaussianSEE}
\end{figure}

What follows are the results for when the original mode $A$ has some
Gaussian weighting functions. We fixed the mass of the scalar field $m=1$.
The metric parameters $a$ and $b$ that determine the initial and final size
of the universe were fixed to $a-b=0.5$ and $a+b=2.5$. In addition, we consider
the case in which $y_{A}(x)=z_{A}(x)=0$, that is no cross terms in $m_{A}$
appear. In figure \ref{GaussianA}, we show the mode $A$ weighting
functions $X_{A}(x)$ and $W_{A}(x)$ after the symplectic transformation. In
figure \ref{GaussianBrho0}, we show the results for partner $B$ weighting
functions $X_{B}(x)$ and $W_{B}(x)$ for the case in which there is no
expansion $\rho =0$. In figure \ref{GaussianBrho10}, we show the results
for partner $B$ weighting functions $X_{B}(x)$ and $W_{B}(x)$ for the case
in which the expansion rate $\rho =10$. Comparing with figure \ref
{GaussianBrho0}, a change of not only the amplitude of the weighting
functions, but also in the width of the functions can be appreciated. As
expected, the partner form is affected by the expansion rate. 
Finally, the entanglement entropy
between mode $A$ and partner $B$ is shown in figure \ref{GaussianSEE} as a
function of the expansion rate $\rho $. It can be seen that the amount of
entanglement between the modes tends to saturate for higher values of the
universe expansion rate $\rho $. The reason is simple. For a large $\rho $,
the scale factor is approximated by a step functional one as
\begin{equation*}
\sqrt{a+b\tanh \left( \rho \eta \right) }=\sqrt{a-b+2b\Theta \left( \eta
\right) }+O(\exp \left( -\rho |\eta |\right) ),
\end{equation*}
where $\Theta \left( \eta\right)$ denotes the step function. 
The metric itself maintains an exponentially small amount of the information
about $\rho $. Hence the entanglement of $A$ and $B$ cannot have high
sensitivity of $\rho $ in this regime. Nevertheless, the entanglement between $A$
and $B$ stores the information of $\rho $.

\begin{figure}[]
\centering
\includegraphics[scale=1]{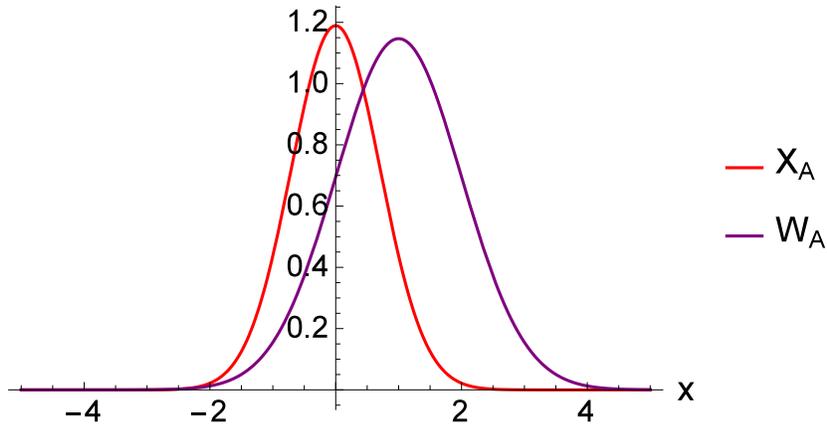}
\caption{Original mode $A$ with Gaussian weighting functions. The weighting
functions $X_A(x)$ and $W_A(x)$ are obtained from the symplectic
transformation of $x_A(x)=e^{-x^2}$ and $w_A(x)=\protect\sqrt{\frac{3}{2\protect\pi}}e^{1/3}e^{-(x-1)^2/2}$, where these functions satisfy the
constraint coming from the canonical commutation relationship. For
simplicity $y_A(x)=0$ and $z_A(x)=0$. }
\label{GaussianA}
\end{figure}

\section{Summary and Discussion}\label{sec_SandD} 
In this paper, we proposed a correlation function
definition of purification partner in Section \ref{sec_definition} for a
given particle mode $A$ in an arbitrary state. This may be useful for
verification experiments of the partner mode. We have also shown the
existence of the partner for arbitrary mode $A$ of a lattice field in a general
state.  For a general Gaussian state, the condition which identifies the
partner is simplified. The entanglement entropy between the mode $A$ and its
partner $B$ is evaluated by using Eqs.~(\ref{eq_g_det}) and (\ref{eq_EE}).
We showed the formula in Eqs.~(\ref{eq_partner_wfs1})-(\ref{eq_partner_wfs2}) to obtain the partner in an arbitrary Gaussian state of
scalar field theory. In addition, we provided a new class of partner:
spatially overlapped partner (SOP). 
As is shown explicitly in an expanding universe
model, the weighting functions of the partner contains information on the
Bogoliubov coefficients, i.e., the partners play a role of a storage of
dynamics information.

As a future work, it is interesting to investigate the advantage of the
identification for SOPs, especially in the context of the black hole
information loss and the cosmological Bell inequality breaking in cosmic
microwave background. As is presented in \cite{TYH}, the purification partners help to enhance the efficiency of entanglement harvesting.

\begin{acknowledgments}
We would like to thank Achim Kempf, Ralf Sch\"utzhold,
Takeshi Tomitsuka, William G. Unruh, and Naoki
Watamura for their useful discussions. This research was partially supported by JSPS KAKENHI Grant Numbers 16K05311 (M.H.) and 18J20057 (K.Y.), and by Graduate Program on Physics for the Universe of Tohoku University (K.Y.).
\end{acknowledgments}

\appendix

\section{\protect\bigskip Proof of Hermitianity, Non-negativity, and
Normalization of $\hat{\protect\rho}_{A}$}\label{append_tilderho}
Let us confirm that $\hat{\rho}_{A}$ is a quantum state, that is, a unit
trace positive-semidefinite Hermitian operator. Since $\chi \left(
x_{A},v_{A}\right) ^{\ast }=\chi \left( -x_{A},-v_{A},\right) $ holds, $\langle \bar{x}_{A}|\hat{\rho}_{A}|x_{A}\rangle ^{\ast }$ is computed as 
\begin{align}
& \langle x_{A}|\hat{\rho}_{A}|\bar{x}_{A}\rangle ^{\ast }  \notag \\
& =\frac{1}{(2\pi )^{2}}\int dv_{A}dv_{B}\chi \left( -(x_{A}-\bar{x}_{A}),-v_{A}\right) e^{+\frac{i}{2}v_{A}\left( \bar{x}_{A}+x_{A}\right) } 
\notag \\
& =\frac{1}{(2\pi )^{2}}\int dv_{A}dv_{B}\chi \left( -(x_{A}-\bar{x}_{A}),v_{A}\right) e^{-\frac{i}{2}v_{A}\left( \bar{x}_{A}+x_{A}\right) } 
\notag \\
& =\langle \bar{x}_{A}|\hat{\rho}_{A}|x_{A}\rangle .
\end{align}
Here, we have changed the sign of integration variables $v_{A}$. Thus, $\hat{\rho}_{A}$ is a Hermitian operator. By using $\langle \Psi |\Psi \rangle =1$, the normalization condition of $\hat{\rho}_{A}$ is directly checked as
follows: 
\begin{align}
\mathrm{Tr}\left[ \hat{\rho}_{A}\right] & =\frac{1}{(2\pi )^{2}}\int
dx_{A}dv_{A}\chi \left( 0,v_{A}\right) e^{-iv_{A}x_{A}}  \notag \\
& =\frac{1}{(2\pi )^{2}}\int dv_{A}\mathrm{Tr}\left( \hat{\rho}e^{iv_{A}\hat{q}_{A}}\right) \int dx_{A}e^{-iv_{A}x_{A}}  \notag \\
& =\int dv_{A}\langle \Psi |e^{iv_{A}\hat{q}_{A}}e^{iv_{B}\hat{q}_{B}}|\Psi
\rangle \delta (v_{A})\delta (v_{B})=\langle \Psi |\Psi \rangle =1.
\end{align}
The operator $\hat{\rho}_{A}$ is positive-semidefinite if and only if 
\begin{equation*}
\int d\bar{x}_{A}dx_{A}\Phi (\bar{x}_{A})^{\ast }\langle \bar{x}_{A}|\hat{\rho}_{A}|x_{A}\rangle \Phi (x_{A})\geq 0
\end{equation*}
holds for any complex function $\Phi (x_{A})$. Let us confirm this
inequality. Substituting the definition of $\langle \bar{x}_{A}|\hat{\rho}_{A}|x_{A}\rangle $ into the above equation, we get 
\begin{align}
& \int d\bar{x}_{A}dx_{A}\Phi (\bar{x}_{A})^{\ast }\langle \bar{x}_{A}|\hat{\rho}_{A}|x_{A}\rangle \Phi (x_{A})  \notag \\
& =\frac{1}{(2\pi )^{2}}\int d\bar{x}_{A}dx_{A}dv_{A}\Phi (\bar{x}_{A})^{\ast }\Phi (x_{A})  \notag \\
& \qquad \times \chi \left( x_{A}-\bar{x}_{A},v_{A}\right) e^{-\frac{i}{2}v_{A}(\bar{x}_{A}+x_{A})}  \notag \\
& =\int d\bar{x}_{A}dx_{A}\Phi (\bar{x}_{A})^{\ast }\Phi (x_{A})  \notag \\
& \qquad \times \langle \Psi |\int \frac{dv_{A}}{2\pi }e^{i\left( v_{A}(\hat{q}_{A}-x_{A})\right) }e^{-i\left( (x_{A}-\bar{x}_{A})\hat{p}_{A}\right)
}|\Psi \rangle ,
\end{align}
where we have used the Baker--Campbell--Hausdorff formula. By using 
\begin{equation*}
\int \frac{dv}{2\pi }e^{i\left( v(\hat{q}-x)\right) }=\delta \left( \hat{q}-x\right) ,
\end{equation*}
and the spectrum decomposition of $\hat{q}_{A}$ : 
\begin{equation*}
\hat{q}_{A}=\sum_{\alpha }\int dx_{A}^{\prime }\,x_{A}^{\prime }\Ket{x_A',\alpha}\Bra{x_A',\alpha},
\end{equation*}
we get 
\begin{align}
& \int d\bar{x}_{A}dx_{A}\Phi (\bar{x}_{A})^{\ast }\langle \bar{x}_{A}|\hat{\rho}_{A}|x_{A}\rangle \Phi (x_{A})  \notag \\
& =\sum_{\alpha ,\beta }\int d\bar{x}_{A}dx_{A}\Phi (\bar{x}_{A})^{\ast
}\Phi (x_{A})  \notag \\
& \qquad \times \Braket{x_A,\alpha|\Braket{x_B,\beta|e^{-i\left((x_A-\bar{x}_A)\hat{p}_A\right)}e^{-i\left((x_B-\bar{x}_B)\hat{p}_B\right)}\hat{\rho}|x_A,\alpha}|x_B,\beta}.
\end{align}
Since $\Bra{x_A,\alpha}e^{-i\left( (x_{A}-\bar{x}_{A})\hat{p}_{A}\right) }=\Bra{\bar{x}_A,\alpha}$ holds, the positive-semidefiniteness is finally
proven as follows: 
\begin{align}
& \int d\bar{x}_{A}dx_{A}\Phi (\bar{x}_{A})^{\ast }\langle \bar{x}_{A}|\hat{\rho}_{A}|x_{A}\rangle \Phi (x_{A})  \notag \\
& =\sum_{\alpha ,\beta }\int d\bar{x}_{A}dx_{A}\Phi (\bar{x}_{A})^{\ast }\Braket{\bar{x}_A,\alpha|\Braket{\bar{x}_B,\beta|\hat{\rho}|x_A,\alpha}|x_B,\beta}\Phi (x_{A})\geq 0.
\end{align}
Therefore, $\hat{\rho}_{A}$ is a quantum state.

\bigskip

\section{Covariance matrix and its standard form}

\label{covariance}

Let us consider a system composed of $N(\geq 2)$ harmonic oscillators whose
canonical variables are given by $(\hat{q}_n,\hat{p}_n)$ for $n=1,\cdots, N$. By using 
\begin{align}
\hat{\bm{r}}\equiv \left(\hat{q}_1,\hat{p}_1,\hat{q}_2,\hat{p}_2,\cdots,\hat{q}_N,\hat{p}_N\right),
\end{align}
the commutation relationships are expressed as 
\begin{align}
\left[\hat{r}_\alpha,\hat{r}_\beta\right]=i\Omega_{\alpha\beta},
\end{align}
where $\Omega$ is defined by 
\begin{align}
\Omega\equiv \bigoplus_{n=1}^N 
\begin{pmatrix}
0 & 1 \\ 
-1 & 0
\end{pmatrix}. \label{eq_j}
\end{align}

A Gaussian state $\hat{\rho}$ is fully characterized by the first and second
moments of canonical variables. By locally shifting the canonical variable,
it is always possible to make the first moments zero. Then, the state $\hat{\rho}$ is characterized by its $2N\times 2N$ covariance matrix: 
\begin{align}
M\equiv \mathrm{Re}\left(\Braket{\hat{\bm{r}}\hat{\bm{r}}\sps{T}}\right),
\end{align}
where $\Braket{\hat{\mathcal{O}}}\equiv \mathrm{Tr}\left(\hat{\rho}\hat{\mathcal{O}}\right)$ denotes the expectation value for a linear operator $\hat{\mathcal{O}}$. It should be noted that the reduced state for $n(<N)$
harmonic oscillators degree of freedom is also Gaussian when the total
system is in a Gaussian state. Thus, for example, the reduced state for a
subsystem composed of the first and the second harmonic oscillators is fully
characterized by its covariance matrix defined by 
\begin{align}
m_{12}\equiv 
\begin{pmatrix}
\Braket{\hat{q}_1^2} & \mathrm{Re}\left(\Braket{\hat{q}_1\hat{p}_1}\right) & \Braket{\hat{q}_1\hat{q}_2} & \Braket{\hat{q}_1\hat{p}_2} \\ 
\mathrm{Re}\left(\Braket{\hat{p}_1\hat{q}_2}\right) & \Braket{\hat{p}_1^2} &
\Braket{\hat{p}_1\hat{q}_2} & \Braket{\hat{p}_1\hat{p}_2} \\ 
\Braket{\hat{q}_2\hat{q}_1} & \Braket{\hat{q}_2\hat{p}_1} & \Braket{\hat{q}_2^2} & \mathrm{Re}\left(\Braket{\hat{q}_2\hat{p}_2}\right)
\\ 
\Braket{\hat{p}_2\hat{q}_1} & \Braket{\hat{p}_2\hat{p}_1} & \mathrm{Re}\left(\Braket{\hat{p}_2\hat{q}_2}\right) & \Braket{\hat{p}_2^2}
\end{pmatrix}
.
\end{align}
In other words, by using the covariance matrix $m_{12}$, we can calculate
the expectation value of any local operator composed of a product of $(\hat{q}_1,\hat{p}_1,\hat{q}_2,\hat{p}_2)$.

A linear transformation $S$ on the canonical variables $\hat{\bm{r}}$ is
called symplectic if and only if $\hat{\bm{R}}\equiv S\hat{\bm{r}}$
satisfies the canonical commutation relationships. This condition is
equivalent to $S\Omega S^{\mathrm{T}} =\Omega$. The Gaussian state $\hat{\rho}$ is also
characterized by the covariance matrix $M^{\prime }$ for the new variable $\hat{\bm{R}}$, which is related with the original one via $M^{\prime }=SMS^{\mathrm{T}} $.

Consider a local symplectic transformation $S$ in the form of 
\begin{align}
S= 
\begin{pmatrix}
S_{1} & 0 & 0 \\ 
0 & S_{2} & 0 \\ 
0 & 0 & \mathbb{I}_{2(N-2)}
\end{pmatrix}
,
\end{align}
where $S_{1}$ and $S_{2}$ are $2\times 2$ symplectic matrices, and $\mathbb{I}_{2(N-2)}$ is the $2(N-2)\times 2(N-2)$ identity matrix. By using this
local transformation, it is known that the covariance matrix $m_{12}$
transforms into the following standard form \cite{S,DGCZ}: 
\begin{align}
M_{12}&\equiv 
\begin{pmatrix}
\Braket{\hat{Q}_1^2} & \mathrm{Re}\left(\Braket{\hat{Q}_1\hat{P}_1}\right) &\Braket{\hat{Q}_1\hat{Q}_2} & \Braket{\hat{Q}_1\hat{P}_2} \\ 
\mathrm{Re}\left(\Braket{\hat{P}_1\hat{Q}_2}\right) & \Braket{\hat{P}_1^2} &
\Braket{\hat{P}_1\hat{Q}_2} & \Braket{\hat{P}_1\hat{P}_2} \\ 
\Braket{\hat{Q}_2\hat{Q}_1} & \Braket{\hat{Q}_2\hat{P}_1} & \Braket{\hat{Q}_2^2} & \mathrm{Re}\left(\Braket{\hat{Q}_2\hat{P}_2}\right)
\\ 
\Braket{\hat{P}_2\hat{Q}_1} & \Braket{\hat{P}_2\hat{P}_1} & \mathrm{Re}\left(\Braket{\hat{P}_2\hat{Q}_2}\right) & \Braket{\hat{P}_2^2}
\end{pmatrix}
\notag \\
&= 
\begin{pmatrix}
a & 0 & c_+ & 0 \\ 
0 & a & 0 & c_- \\ 
c_+ & 0 & b & 0 \\ 
0 & c_- & 0 & b
\end{pmatrix}
,
\end{align}
where 
\begin{align}
\begin{pmatrix}
\hat{Q}_i \\ 
\hat{P}_i
\end{pmatrix}
\equiv S_i 
\begin{pmatrix}
\hat{q}_i \\ 
\hat{p}_i
\end{pmatrix}
\end{align}
for $i=1,2$ and $a\geq 0$, $b\geq 0$ and $c_\pm\in\mathbb{R}$. The reduced
state for the subsystem composed of the first and second oscillator is pure
if and only if 
\begin{align}
a=b, \quad c_+=-c_-,\quad c_+c_-=\frac{1}{4}-a^2
\end{align}
hold \cite{S,DGCZ}. Therefore, the second harmonic oscillator purifies the
first one if 
\begin{align}
M_{12}= 
\begin{pmatrix}
\frac{1}{2}\sqrt{1+g^2} & 0 & \frac{g}{2} & 0 \\ 
0 & \frac{1}{2}\sqrt{1+g^2} & 0 & -\frac{g}{2} \\ 
\frac{g}{2} & 0 & \frac{1}{2}\sqrt{1+g^2} & 0 \\ 
0 & -\frac{g}{2} & 0 & \frac{1}{2}\sqrt{1+g^2}
\end{pmatrix}
\end{align}
holds, where $g$ is a positive number. This condition plays a crucial role
to obtain the partner formula. The factor $g$ is directly related with the
entanglement entropy $S_{\mathrm{EE}} $ between the first and second
harmonic oscillator as follows \cite{HSH}: 
\begin{align}
S_{\mathrm{EE}} =\sqrt{1+g^2}\ln{\left(\frac{1}{g}\left(\sqrt{1+g^2}+1\right)\right)}+\ln{\left(\frac{g}{2}\right)}.
\end{align}

\end{document}